\documentclass[aps,prb,reprint,superscriptaddress]{revtex4-2}

\usepackage{amsmath}
\usepackage{amsthm}
\usepackage{blindtext}
\usepackage{bm}
\usepackage{amsfonts}
\usepackage{graphicx}
\usepackage{amssymb}
\usepackage{feynmp}
\usepackage{breqn}
\usepackage{stackengine}
\usepackage{diagbox}
\usepackage{booktabs}
\usepackage{tikz}
\usepackage{tikz-feynman}
\usepackage[normalem]{ulem}

\newcommand{\ket}[1]{| #1 \rangle}

\newcommand{\avg}[1]{\langle #1 \rangle}

\newcommand{\kvec}{\mathbf{k}}
\newcommand{\qvec}{\mathbf{q}}

\newcommand{\Hop}{\mathcal{H}}

\newcommand{\Uop}{\mathcal{U}}

\makeatletter
\let\cat@comma@active\@empty
\makeatother

\begin{document}


\title{Excitonic correlations in the equilibrium and voltage-biased bilayer Hubbard model: multi-orbital two-particle self-consistent approach}


\author{Jiawei Yan}
\affiliation{2020 X-Lab, Shanghai Institute of Microsystem and Information Technology, \\
Chinese Academy of Sciences, Shanghai 200050, China}
\affiliation{Department of Physics, University of Fribourg, 1700 Fribourg, Switzerland}
\author{Jonas B.~Profe}
\affiliation{Institute for Theoretical Physics, Goethe University Frankfurt,
Max-von-Laue-Straße 1, D-60438 Frankfurt a.M., Germany}
\author{Yuta Murakami}
\affiliation{Center for Emergent Matter Science, RIKEN, Saitama 351-0198, Japan}
\affiliation{Institute for Materials Research, Tohoku University, Sendai 980-8577, Japan}
\author{Philipp Werner}
\affiliation{Department of Physics, University of Fribourg, 1700 Fribourg, Switzerland}


\date{\today}

\begin{abstract}
We develop a nonequilibrium multi-orbital extension of the two-particle self-consistent theory and apply it to the bilayer Hubbard model as a minimal platform to investigate correlation effects in the presence of interlayer interactions and tunneling.
The method determines vertex corrections in the spin and charge channels self-consistently at the two-particle level, thereby avoiding the spurious finite-temperature phase transitions that limit dynamical mean-field theory in two dimensions.
We derive the spectral self-energy and implement the framework directly on the real-frequency axis within the Keldysh nonequilibrium Green’s function formalism, enabling the treatment of both equilibrium and non-equilibrium steady states without relying on numerical analytic continuation.
As an application, we demonstrate that a pseudogap can emerge in the bilayer Hubbard model when spin, charge, or excitonic fluctuations become sufficiently strong.
Instabilities in different channels are also evaluated in an unbiased manner across the parameter space.
Remarkably, we find that the excitonic susceptibility grows with increasing interlayer bias, before it gets suppressed at large biases by the charge imbalance between the layers.
This work establishes a versatile and computationally efficient framework for investigating correlated multi-orbital systems under nonequilibrium conditions.
\end{abstract}


\hyphenation{single}

\maketitle

\section{Introduction\label{sec: introduction}}

Understanding the emergence of collective phenomena in correlated electron systems remains one of the central challenges in modern condensed matter physics.
Strong electron–electron interactions often give rise to magnetism, superconductivity, charge-density-wave order, exciton condensates, and pseudogap behavior, particularly in low-dimensional materials where quantum fluctuations play a crucial role \cite{Dagotto2005,Bruus2004,Coleman2015,Altland2010,Lee2006}.
This challenge becomes even more pronounced in multi-orbital systems, where crystal field splitting, Hund’s coupling, and inter-orbital hybridization introduce additional channels for competing instabilities \cite{Georges2013,Kunes_2015,Hoshino2016,Galler2017}.
If the electron populations in correlated systems are driven out of equilibrium by the application of voltage biases or laser pulses, the stability of the various phases is modified and new nonthermal orders can emerge \cite{Stefanucci2013,Eisert2015,Aoki2014,Murakami2025}.
With the advent of engineered heterostructures, twisted bilayers, and bilayer transition-metal-based compounds, there is an increasing demand for theoretical frameworks capable of capturing both multi-orbital correlations and nonequilibrium dynamics.

A variety of theoretical approaches have been developed to study correlated lattice systems.
Among them, dynamical mean-field theory (DMFT) \cite{Georges1996} and its cluster extensions \cite{Maier2005} have become standard tools for analyzing local correlations and Mott physics.
However, DMFT is intrinsically a local theory, and when applied to low-dimensional systems, it tends to produce spurious symmetry-breaking phase transitions at nonzero temperature due to the neglect of nonlocal fluctuations \cite{Schaefer2015,Schaefer2021}.
Diagrammatic approaches such as the fluctuation exchange (FLEX) approximation \cite{Bickers1989} can incorporate both local and nonlocal fluctuations perturbatively.
However, the fluctuations included in FLEX do not take into account vertex corrections, which 
leads to a violation of 
the Ward identities \cite{Baym1962,*Baym1961}. 
As a result, the obtained two-particle response functions are not norm-conserving and become unreliable in the strongly correlated regime.

The two-particle self-consistent (TPSC) approach \cite{Vilk1997b,Allen2004,Tremblay2011} offers a complementary perspective.
By enforcing exact local sum rules on two-particle correlation functions, TPSC provides a self-consistent framework for incorporating vertex corrections in both the spin and charge channels, thereby avoiding the artificial ordering tendencies that plague mean-field methods in two dimensions.
Over the past two decades, TPSC has been successfully applied to single-band Hubbard models, capturing pseudogap behavior, spin fluctuations, and superconducting instabilities in parameter regimes where controlled methods are scarce \cite{Moukouri2000,Allen2001,Arya2015,Tremblay2006}.
While most TPSC applications have been limited to equilibrium and single-orbital systems,
there have been recent efforts to extend the TPSC to multi-orbital Hubbard models with Kanamori-type interactions \cite{Profe2025,andp.202000399,Zantout2019,Miyahara2013}.
In the meantime, nonequilibrium Green's function theory has also been employed to extend the single-band theory to nonequilibrium conditions for both time-dependent dynamics and steady-state simulations \cite{PhysRevB.106.L241110,PhysRevB.109.155113,PhysRevB.105.085122}.
These advances lay the foundation for a combined nonequilibrium multi-orbital TPSC, which so far has not been realized, but can become a promising framework for simulating low-dimensional materials under nonequilibrium conditions.

On the physics side, recent years have witnessed significant progress in understanding excitonic instabilities in multi-orbital and bilayer Hubbard models.
Theoretical studies showed that interlayer Coulomb interactions can drive the formation of bound electron–hole pairs (excitons) across the layers, leading to the possibility of excitonic condensation and novel quantum phases \cite{Kunes_2015,Rademaker2013,Kaneko2013,Su2017}.
Advanced numerical techniques, such as DMFT \cite{Giuli2023}, determinant quantum Monte Carlo \cite{PhysRevB.88.235115}, and functional renormalization group (fRG) approaches \cite{Scherer2018} have been employed to map out the phase diagram and characterize the competition between excitonic order, magnetism, and charge fluctuations.
Particular attention has also been devoted to the BCS-BEC crossover in the excitonic phase, which captures the smooth evolution from weakly bound, Cooper-like electron–hole pairs in a semimetal to tightly bound excitons characteristic of a semiconductor at very low temperatures \cite{Zenker2011,Zenker2012}.
These developments established the bilayer Hubbard model as a minimal and versatile platform for exploring excitonic phenomena in correlated electron systems.

In this work, we present a nonequilibrium multi-orbital TPSC approach and apply it to the bilayer Hubbard model.
This bilayer setup serves as an ideal testbed for exploring how on-site and interlayer interactions and hoppings modify spin and charge responses, while simultaneously assessing the robustness of multi-orbital TPSC beyond the single-band case.
We employ a Hartree-Fock ansatz, recently proposed in Ref.~\cite{Profe2025}, to compute the renormalized vertices.
Our formalism is implemented directly on the real-frequency axis using Keldysh nonequilibrium Green's function theory for steady-state calculations, which circumvents the need for numerically ill-defined analytical continuation \cite{Keldysh1965}.
The results demonstrate that within the bilayer model, both on-site and interlayer spin vertices are effectively screened, which prevents spurious finite-temperature phase transitions in two dimensions.
Nonetheless, at low temperatures, a pseudogap emerges close to the Fermi level, signaling a precursor to the transition.
By comparing susceptibilities across different channels, we construct a phase diagram that illustrates the competition between the on-site interaction $U$, interlayer interaction $V$, and crystal field splitting $\Delta \epsilon$.
We furthermore demonstrate that an intermediate voltage bias perpendicular to the layers can stabilize excitonic bound states; however, this enhancement is eventually suppressed by the charge imbalance between the layers.

The rest of the paper is organized as follows.
In Sec.~\ref{sec: formalism}, we introduce the model Hamiltonian and outline the general formalism, including many-body perturbation theory, the TPSC vertex renormalization scheme, and the derivation of the spectral self-energy.
Section~\ref{sec: numerical results} presents the numerical results for the bilayer Hubbard model.
Specifically, Sec.~\ref{subsec: order parameter} defines the order parameters and the associated susceptibilities, while Sec.~\ref{subsec: rpa} provides a preliminary RPA analysis of the model.
The TPSC results are reported in Secs.~\ref{subsec: tpsc for wperp = 0} and \ref{subsec: tpsc for wperp != 0} for both equilibrium and nonequilibrium situations.
Finally, we summarize our findings in Sec.~\ref{sec: conclusions}.

\section{Formalism\label{sec: formalism}}

\subsection{Model Hamiltonian\label{subsec: model hamiltonian}}

We discuss the two-particle self-consistent approach for 
a generic multi-orbital Hubbard model with Hamiltonian $\Hop(t) = \Hop^0(t) + \Hop^{int}(t)$, where the non-interacting and the interacting parts take the form
\begin{subequations}\label{eq: Hamiltonian}
\begin{align}
\Hop^0(t) &= 
T_{\bar{i}\bar{j},\bar{\alpha}\bar{\beta}}(t) c_{\bar{i}\bar{\alpha}\bar{\sigma}}^\dag c_{\bar{j}\bar{\beta}\bar{\sigma}}~,\label{eq: non-interacting Hamiltonian}\\
\Hop^{int}(t) &= \frac{1}{4} 
U_{\bar{i},\bar{\alpha}\bar{\beta}\bar{\gamma}\bar{\delta},\bar{\sigma}_1\bar{\sigma}_2\bar{\sigma}_3\bar{\sigma}_4}(t) c_{\bar{i}\bar{\alpha}\bar{\sigma}_1}^\dag c_{\bar{i}\bar{\beta}\bar{\sigma}_2}^\dag c_{\bar{i}\bar{\delta}\bar{\sigma}_4} c_{\bar{i}\bar{\gamma}\bar{\sigma}_3}~.\label{eq: interacting Hamiltonian}
\end{align}
\end{subequations}
Here, the site indices are denoted by $i$ and $j$, the orbital indices by $\alpha$ through $\delta$, and the spin indices by $\sigma_1$ through $\sigma_4$, respectively.
In the following, indices with overbars denote dummy variables that are implicitly summed over, while those without overbars represent external indices.
$T_{ij,\alpha\beta}$ represents the kinetic energy matrix, with the chemical potential absorbed into the on-site energy.
Due to the SU(2) spin symmetry, the Coulomb tensor takes an asymmetric form in the spin space
\begin{dmath}
U_{i,\alpha\beta\gamma\delta,\sigma_1\sigma_2\sigma_3\sigma_4} = U_{i,\alpha\beta\gamma\delta}\delta_{\sigma_1\sigma_3}\delta_{\sigma_2\sigma_4} - U_{i,\alpha\beta\delta\gamma}\delta_{\sigma_1\sigma_4}\delta_{\sigma_2\sigma_3}~.
\end{dmath}
For Kanamori-type interactions \cite{Georges2013}, 
\begin{dmath}\label{eq: Kanamori-type interaction}
U_{i,\alpha\beta\gamma\delta} = U_{i,\alpha\beta}\delta_{\alpha\gamma}\delta_{\beta\delta} + J_{i,\alpha\beta} \delta_{\alpha\delta}\delta_{\beta\gamma} + J^C_{i,\alpha\gamma}\delta_{\alpha\beta}\delta_{\gamma\delta}~,
\end{dmath}
where $U_{\alpha\beta}$ and $J^{(C)}_{\alpha\beta}$ are the Coulomb and Hund's couplings between different orbitals.
In general, the parameters $T$ and $U$ in Eq.~\eqref{eq: Hamiltonian} can be time dependent.

The TPSC approach is based on renormalized many-body perturbation theory and was originally formulated for the single-orbital Hubbard model \cite{Vilk1994a,Vilk1994,Vilk1997b}.
It has recently been extended to multi-orbital setups \cite{Profe2025} and here we present a real-frequency implementation which can also be used to study nonequilibrium problems.
Within the nonequilibrium Green's function (NEGF) theory \cite{Stefanucci2013}, the temporal variables in Eq.~\eqref{eq: Hamiltonian} are generalized from real times $t$ to contour times $z$.
To simplify the notation in the subsequent discussion, we introduce a combined index $1=(i, z, \cdots)$ to represent all indices (site, orbital, time, spin, etc.) that are not explicitly specified in the subscripts.

\subsection{Basics of many-body perturbation theory\label{subsec: mbpt}}

To derive the theory, we start with the equation of motion of the single-particle (1P) Green's function (GF)
\begin{dmath}
G_{\alpha\beta,\sigma\sigma'}(1,2) = -i \avg{ \mathbb{T} c_{\alpha\sigma}(1) c^\dag_{\beta\sigma'}(2) }~.
\end{dmath}
By taking the time derivative of $G$, and using the Dyson equation
\begin{dmath}
[G^{-1}](1,2) = [(G^0)^{-1}](1,2) - \Sigma(1,2)~,
\end{dmath} 
we obtain the self-energy expression
\begin{dmath}\label{eq: equation of motion}
[\Sigma * G]_{\alpha\beta,\sigma\sigma'}(1,2) = -i U_{\alpha\bar{\gamma}\bar{\delta}\bar{\epsilon}}(1) G_{\bar{\delta}\bar{\epsilon}\beta\bar{\gamma},\sigma\bar{\sigma}\sigma'\bar{\sigma}}(1,1,2,1^+)~,
\end{dmath}
where $G$ on the r.h.s. is the two-particle (2P) Green's function \footnote{We use the same symbol as it can be easily distinguished from the 1P GF by the number of the indices or arguments.}
\begin{dmath}
G(1,2,3,4) = (-i)^2 \avg{\mathbb{T} c(1) c(2) c^\dag(4) c^\dag(3) }~.
\end{dmath}
Here, $\mathbb{T}\{\cdots\}$ orders the operators from right to left with increasing contour time $z$.
$G^0$ is the non-interacting Green's function corresponding to $\Hop^0$ in Eq.~\eqref{eq: Hamiltonian}.
The notation $1^+$ in Eq.~\eqref{eq: equation of motion} indicates that the temporal variable is shifted by an infinitesimal time along the contour.

We next introduce the 4-point response function $\chi(1,2,3,4)$, which is defined as the response of the 1P GF under an external field in the particle-hole (ph) channel, i.e. $i\chi(1,2,3,4) = \delta G(1,3) / \delta \phi(4,2) | _{\phi = 0}$ \cite{Kadanoff1962}.
From this definition, $\chi$ is related to the 2P GF by 
\begin{dmath}\label{eq: generalized susceptibility}
i\chi(1,2,3,4) = G(1,3)G(2,4) - G(1,2,3,4)~,
\end{dmath}
and it obeys a Dyson-like equation on the 2P level, known as the Bethe-Salpeter equation (BSE):
\begin{dmath}\label{eq: bethe-salpeter equation}
i\chi(1,2,3,4) = G(1,4)G(2,3) - iG(1,\bar{1}) G(\bar{3},3) \Lambda(\bar{1},\bar{2},\bar{3},\bar{4}) \chi(\bar{4},2,\bar{2},4)~.
\end{dmath}
Note that $G(1,3)G(2,4)$ in Eq.~\eqref{eq: generalized susceptibility} is canceled by the direct propagation of $G(1,2,3,4)$, leaving only the exchange propagation $G(1,4)G(2,3)$ in Eq.~\eqref{eq: bethe-salpeter equation}.
$\Lambda(1,2,3,4) = -\delta \Sigma(1,3) / \delta G(4,2)$ in Eq.~\eqref{eq: bethe-salpeter equation} is the 2P irreducible vertex, which contains all the 2P irreducible diagrams in the ph-channel.
However, $\Lambda$ depends on four external indices, making it computationally challenging to handle this object in multi-orbital and nonequilibrium situations.
To deal with this, the simplest solution is the random phase approximation (RPA), which replaces $\Lambda$ with the bare Coulomb interaction $U$.
However, conventional RPA is only valid in the weak-coupling regime, due to the rapid divergence of the susceptibility in the spin channel.
To address this issue, the TPSC approach assumes that the irreducible vertex function $\Lambda$ in Eq.~\eqref{eq: bethe-salpeter equation} is local in both time and space,  like RPA.
However, its value is renormalized in both the spin and charge channels individually, i.e.
\begin{dmath}\label{eq: local vertex approximation}
\Lambda^{sp/ch}_{\alpha\beta\gamma\delta}(1,2,3,4) = i\tilde{\Lambda}^{sp/ch}_{\alpha\beta\gamma\delta}(1) \delta(1-3) \delta(1-4) \delta(1^+ - 2)~.
\end{dmath}
The renormalized vertices $\tilde{\Lambda}^{sp/ch}$, in TPSC, are determined self-consistently on the 2P level, as will be discussed in the next subsection.
In Eq.~\eqref{eq: local vertex approximation}, the $\Lambda$s are transformed into spin and charge sectors according to $\Lambda^{sp/ch} = \Lambda_{\uparrow\downarrow\uparrow\downarrow} \mp \Lambda_{\uparrow\uparrow\uparrow\uparrow}$.
This transformation is valid since we restrict our calculations to paramagnetic solutions, making the 1P quantities spin independent and the 2P quantities dependent on two spin indices.
The TPSC Ansatz significantly simplifies solving the BSEs and facilitates their extension to non-equilibrium problems.
It is worth noting that, in contrast to the Hartree decomposition used in the previous literature \cite{Zantout2019}, we employ the Hartree-Fock Ansatz, which provides substantial improvements over the former scheme for small Hund's coupling $J$ \cite{Profe2025}.
By further introducing the spin and charge response functions $\chi^{sp/ch} = 2 \left( \chi_{\uparrow\uparrow\uparrow\uparrow} \mp \chi_{\uparrow\downarrow\uparrow\downarrow} \right)$, Eq.~\eqref{eq: bethe-salpeter equation} under the chosen Ansatz reduces to
\begin{dmath}\label{eq: bethe-salpeter equation simplified}
\chi^{sp/ch}_{ab}(1,2) = \chi^0_{ab}(1,2) \mp \frac{1}{2} \chi^0_{a\bar{a}}(1,\bar{1}) \tilde{\Lambda}^{sp/ch}_{\bar{a}\bar{b}}(\bar{1}) \chi^{sp/ch}_{\bar{b}b}(\bar{1},2)~,
\end{dmath}
where $a = (\alpha\gamma)$ and $b = (\delta\beta)$ are grouped indices, which allow to convert the summation in orbital space into a matrix multiplication (see Appendix \ref{sec: orbital resolved BSE}).
Here, $\chi(1,2) = \chi(1,2,1^+,2^+)$ is the 2-point response function and $\chi^0(1,2) = -iG(1,2^+)G(2,1^+)$ is the bare electron-hole bubble.

\subsection{Renormalization of spin and charge vertices\label{subsec: spin and charge renormalization}}

As mentioned above, the key step in the TPSC theory is to renormalize the spin and charge vertices, as shown in Eq.~\eqref{eq: local vertex approximation}.
To derive explicit expressions, we start with the exact equation of motion, Eq.~\eqref{eq: equation of motion}, and rewrite its r.h.s. by dividing and multiplying the HF decomposition of the 2P GF by the same factor \footnote{The Ansatz is not unique, see Ref.~\cite{andp.202000399}}
\begin{widetext}
\begin{dmath}
[\Sigma * G]_{\alpha\beta,\sigma\sigma'}(1,2) 
= -i
U_{\alpha\bar{\gamma}\bar{\delta}\bar{\epsilon}}(1) \frac{G_{\bar{\delta}\bar{\epsilon}\beta\bar{\gamma},\sigma\bar{\sigma}\sigma'\bar{\sigma}}(1,1,2,1^+)}{G_{\bar{\delta}\beta,\sigma\sigma'}(1,2)G_{\bar{\epsilon}\bar{\gamma},\bar{\sigma}\bar{\sigma}}(1,1^+) - G_{\bar{\delta}\bar{\gamma},\sigma\bar{\sigma}}(1,1^+) G_{\bar{\epsilon}\beta,\bar{\sigma}\sigma'}(1,2) } \\
\times \left[ G_{\bar{\delta}\beta,\sigma\sigma'}(1,2)G_{\bar{\epsilon}\bar{\gamma},\bar{\sigma}\bar{\sigma}}(1,1^+) - G_{\bar{\delta}\bar{\gamma},\sigma\bar{\sigma}}(1,1^+) G_{\bar{\epsilon}\beta,\bar{\sigma}\sigma'}(1,2) \right]~.
\end{dmath}
The TPSC Ansatz introduces a renormalization factor $\lambda$, which is the local part of the fraction in the above equation
\begin{dmath}\label{eq: renormalization factor}
\lambda_{\alpha\beta\gamma\delta,\sigma_1\sigma_2\sigma_3\sigma_4}(1) 
=
\frac{ D_{\alpha\beta\gamma\delta,\sigma_1\sigma_2\sigma_3\sigma_4}(1)}{ n_{\alpha\gamma,\sigma_1\sigma_3}(1) n_{\beta\delta,\sigma_2 \sigma_4}(1) - n_{\alpha\delta,\sigma_1\sigma_4}(1) n_{\beta\gamma,\sigma_2 \sigma_3}(1) }~.
\end{dmath}
Here, we defined a (generalized) double occupancy tensor $D_{\alpha\beta\gamma\delta,\sigma_1\sigma_2\sigma_3\sigma_4}(1) = -G_{\delta\gamma\beta\alpha,\sigma_4\sigma_3\sigma_2\sigma_1}(1,1,1^+,1^+)
= \avg{ c_{\alpha\sigma_1}^\dag(1) c_{\beta\sigma_2}^\dag(1) c_{\delta\sigma_4}(1) c_{\gamma\sigma_3}(1) }$.
As a result, Eq.~\eqref{eq: equation of motion} under the Hartree-Fock Ansatz reduces to 
\begin{dmath}\label{eq: approximated eom}
[\Sigma * G]_{\alpha\beta,\sigma\sigma'}(1,2)
\approx
-i A_{\alpha\bar{\gamma}\bar{\delta}\bar{\epsilon},\sigma\bar{\sigma}\sigma\bar{\sigma}}(1) 
\left[ G_{\bar{\delta}\beta,\sigma\sigma'}(1,2)G_{\bar{\epsilon}\bar{\gamma},\bar{\sigma}\bar{\sigma}}(1,1^+) - G_{\bar{\delta}\bar{\gamma},\sigma\bar{\sigma}}(1,1^+) G_{\bar{\epsilon}\beta,\bar{\sigma}\sigma'}(1,2) \right]~,
\end{dmath}
where we introduced 
$A_{\alpha\beta\gamma\delta,\sigma_1\sigma_2\sigma_3\sigma_4}(1) = U_{\alpha\beta\gamma\delta}(1) \lambda_{\alpha\beta\gamma\delta,\sigma_1\sigma_2\sigma_3\sigma_4}(1)$.
It is easy to verify that when $\beta = \alpha$, $\sigma' = \sigma$ and $2 \rightarrow 1^{+}$, Eq.~\eqref{eq: approximated eom} reduces to the exact equation of motion (Eq.~\eqref{eq: equation of motion}), since this is the way we defined $\lambda$.
The intuition behind this approximation is that the double occupancy tensor $D$ is local in both space and time (but not in the orbital space), and thus not quite sensitive to environmental changes.

By multiplying Eq.~\eqref{eq: approximated eom}  with $G^{-1}$, one obtains the explicit self-energy expression 
\begin{dmath}\label{eq: self-energy with HF Ansatz}
\Sigma_{\alpha\gamma,\sigma_1\sigma_3}(1,3) = -i\delta(1-3)
\left[
\delta_{\sigma_1\sigma_3} A_{\alpha\bar{\eta}\gamma\bar{\epsilon},\sigma_1\bar{\sigma}\sigma_1\bar{\sigma}}(1) G_{\bar{\epsilon}\bar{\eta},\bar{\sigma}\bar{\sigma}}(1,1^+)
- A_{\alpha\bar{\eta}\bar{\epsilon}\gamma,\sigma_1\sigma_3\sigma_1\sigma_3}(1) G_{\bar{\epsilon}\bar{\eta},\sigma_1\sigma_3}(1,1^+)
\right]~.
\end{dmath}
The thermodynamically consistent particle-hole irreducible vertex can be obtained from the functional derivative of $\Sigma$ w.r.t. $G$, i.e. $\Lambda(1,2,3,4) = - \delta \Sigma(1,3) / \delta G(4,2)$.
This leads to the spin part of Eq.~\eqref{eq: local vertex approximation}, namely
\begin{dmath}
\Lambda^{sp}_{\alpha\beta\gamma\delta}(1,2,3,4) = 
i \tilde{\Lambda}^{sp}_{\alpha\beta\gamma\delta}(1)
\delta(1-3) \delta(1-4) \delta(1^+ - 2)~,
\end{dmath}
where $\tilde{\Lambda}^{sp}_{\alpha\beta\gamma\delta}(1) = A_{\alpha\beta\gamma\delta,\uparrow\downarrow\uparrow\downarrow}(1) - A_{\alpha\beta\gamma\delta,\uparrow\uparrow\uparrow\uparrow}(1) + A_{\alpha\beta\delta\gamma,\uparrow\uparrow\uparrow\uparrow}(1)$.
Note that $A$ depends on $G$ when taking the functional derivative of $\Sigma$ in Eq.~\eqref{eq: self-energy with HF Ansatz}.
However, in the SU(2) symmetric case, $\delta A / \delta G$ is exactly canceled out in the spin channel.

Determining the spin vertices is equivalent to calculating $D$, which involves five independent types of local correlation functions.
Specifically, $D_{\alpha\alpha\alpha\alpha,\uparrow\downarrow\uparrow\downarrow}$, $D_{\alpha\beta\alpha\beta,\uparrow\downarrow\uparrow\downarrow}$, $D_{\alpha\beta\beta\alpha,\uparrow\downarrow\uparrow\downarrow}$ and $D_{\alpha\alpha\beta\beta,\uparrow\downarrow\uparrow\downarrow}$  with anti-parallel spin indices, and $D_{\alpha\beta\alpha\beta,\uparrow\uparrow\uparrow\uparrow}$ with parallel spin indices.
In order to uniquely determine them, one has to find five independent equations.
Inspired by the idea of vanilla TPSC, we resort to local spin sum rules.
Specifically, we define a 2-point spin susceptibility as $\chi^{sp}_{\alpha\beta\gamma\delta}(1,2) = \chi^{sp}_{\alpha\beta\gamma\delta}(1,2,1^+,2^+)$.
By setting $2\rightarrow 1^+$, we obtain five exact equalities, also known as local spin sum rules for multi-orbital systems: 
\begin{subequations}\label{eq: spin sum rules}
\begin{align}
\chi_{\alpha\alpha\alpha\alpha}^{sp}(1,1^+) &=  -in_{\alpha\alpha}^t(1) + 2i D_{\alpha\alpha\alpha\alpha,\uparrow\downarrow\uparrow\downarrow}(1)~,\label{eq: spin sum rules -- a}\\
\chi_{\alpha\beta\alpha\beta}^{sp}(1,1^+) &= 2i\left[ D_{\alpha\beta\alpha\beta,\uparrow\downarrow\uparrow\downarrow}(1) - D_{\alpha\beta\alpha\beta,\uparrow\uparrow\uparrow\uparrow}(1)\right] - \delta_{\alpha\beta} in^t_{\alpha\alpha}(1)~, \label{eq: spin sum rules -- b}\\
\chi_{\alpha\beta\beta\alpha}^{sp}(1,1^+) &= -in_{\alpha\alpha}^t(1) + 2i \left[ D_{\alpha\beta\beta\alpha,\uparrow\downarrow\uparrow\downarrow}(1) + D_{\alpha\beta\alpha\beta,\uparrow\uparrow\uparrow\uparrow}(1) \right]
\overset{SU(2)} = -in^t_{\alpha\alpha}(1) + 2i D_{\alpha\beta\alpha\beta,\uparrow\downarrow\uparrow\downarrow}(1)~, \label{eq: spin sum rules -- c}\\
\chi_{\alpha\alpha\beta\beta}^{sp}(1,1^+) &= -\delta_{\alpha\beta} i n_{\alpha\alpha}^t(1) + 2i D_{\beta\beta\alpha\alpha,\uparrow\downarrow\uparrow\downarrow}(1)~.\label{eq: spin sum rules -- d}
\end{align}
\end{subequations}
The second equality of the third equation only holds in the SU(2) symmetric case, since the identity $
\avg{c_\uparrow^\dag(1) c_{\uparrow}^\dag(2) c_\uparrow(4) c_\uparrow(3)} = 
\avg{c_\uparrow^\dag(1) c_{\downarrow}^\dag(2) c_\downarrow(4) c_\uparrow(3)}
+
\avg{c_\uparrow^\dag(1) c_{\downarrow}^\dag(2) c_\uparrow(4) c_\downarrow(3)}
$
has been used to derived it.
These equations are equivalent to the results obtained in Ref.~\cite{Profe2025}.
In the single-band case, only the first sum rule survives and the equations reduce to the conventional TPSC formalism.
With Eq.~\eqref{eq: renormalization factor} and Eq.~\eqref{eq: spin sum rules}, one can self-consistently iterate the spin vertices as well as the generalized double occupancies.
Specifically, an initial value is guessed for the renormalization factor $\lambda$, which sets the spin vertices $\Lambda^{sp}$.
The BSE in the spin channel is then solved to yield the spin susceptibility $\chi^{sp}$ found on the left-hand side of Eq.~\eqref{eq: spin sum rules}.
Next, the double occupancies are computed through the following sequence:
(i) solve for $D_{\alpha\alpha\alpha\alpha,\uparrow\downarrow\uparrow\downarrow}$ using Eq.~\eqref{eq: spin sum rules -- a};
(ii) solve for $D_{\beta\beta\alpha\alpha,\uparrow\downarrow\uparrow\downarrow}$ using Eq.~\eqref{eq: spin sum rules -- d};
(iii) determine $D_{\alpha\beta\alpha\beta,\uparrow\downarrow\uparrow\downarrow}$ from the second equality in Eq.~\eqref{eq: spin sum rules -- c};
(iv) using the result from step (iii), calculate $D_{\alpha\beta\alpha\beta,\uparrow\uparrow\uparrow\uparrow}$ from Eq.~\eqref{eq: spin sum rules -- b};
(v) finally, with the result from step (iv), solve for $D_{\alpha\beta\beta\alpha,\uparrow\downarrow\uparrow\downarrow}$ using the first equality in Eq.~\eqref{eq: spin sum rules -- c}.
These double occupancies are used to update the renormalization factor $\lambda$ via Eq.~\eqref{eq: renormalization factor}. This entire procedure is iterated until the spin vertices converge.

In the charge channel, the vertex is defined by $\Lambda^{ch} = \Lambda_{\uparrow\downarrow\uparrow\downarrow} + \Lambda_{\uparrow\uparrow\uparrow\uparrow}$.
In contrast to the spin channel, where $\frac{\delta A}{\delta G_{\downarrow\downarrow}} - \frac{\delta A}{\delta G_{\uparrow\uparrow}}$ cancels in states without any orders, there is no cancellation of $\frac{\delta A}{\delta G_{\downarrow\downarrow}} + \frac{\delta A}{\delta G_{\uparrow\uparrow}}$ in the charge channel.
For simplicity, we neglect the $\delta A/\delta G$ contributions from the charge channel, and assume that the structure of the charge vertex is the same as that of the spin vertex, 
\begin{dmath}
\Lambda^{ch}_{\alpha\beta\gamma\delta}(1,2,3,4) = i\tilde{\Lambda}^{ch}_{\alpha\beta\gamma\delta}(1) \delta(1-3) \delta(1-4) \delta(1^+ - 2)~.
\end{dmath}
In order to determine $\tilde{\Lambda}^{ch}$, one can similarly introduce local charge sum rules by taking the limit $2 \rightarrow 1^+$ in the 2-point charge susceptibility $\chi^{ch}_{\alpha\beta\gamma\delta}(1,2) = \chi^{ch}_{\alpha\beta\gamma\delta}(1,2,1^+,2^+)$.
As a result, we have
\begin{subequations}\label{eq: charge sum rules}
\begin{align}
\chi_{\alpha\alpha\alpha\alpha}^{ch}(1,1^+) &= i[n_{\alpha\alpha}^t(1)]^2 - in_{\alpha\alpha}^t(1) - 2iD_{\alpha\alpha\alpha\alpha,\uparrow\downarrow\uparrow\downarrow}(1)~,\\
\chi_{\alpha\beta\alpha\beta}^{ch}(1,1^+) &= in_{\alpha\alpha}^t(1)n_{\beta\beta}^t(1) -\delta_{\alpha\beta}in_{\alpha\alpha}^t(1) - 2i\left[ D_{\alpha\beta\alpha\beta,\uparrow\uparrow\uparrow\uparrow}(1) + D_{\alpha\beta\alpha\beta,\uparrow\downarrow\uparrow\downarrow}(1) \right]~,\\
\chi_{\alpha\beta\beta\alpha}^{ch}(1,1^+) &= i \left(n_{\beta\alpha}^t(1) n_{\alpha\beta}^t(1) - n_{\alpha\alpha}^t(1) \right) +2i \left[ D_{\alpha\beta\alpha\beta,\uparrow\uparrow\uparrow\uparrow}(1) - D_{\alpha\beta\beta\alpha,\uparrow\downarrow\uparrow\downarrow}(1) \right]~,\\
&\overset{SU(2)}= in_{\beta\alpha}^t(1) n_{\alpha\beta}^t(1) - in_{\alpha\alpha}^t(1) + 4i D_{\alpha\beta\alpha\beta,\uparrow\uparrow\uparrow\uparrow}(1) - 2i D_{\alpha\beta\alpha\beta,\uparrow\downarrow\uparrow\downarrow}(1)~,\\
\chi_{\alpha\alpha\beta\beta}^{ch}(1,1^+) &= i[n_{\beta\alpha}^t(1)]^2 - \delta_{\alpha\beta} in_{\alpha\alpha}^t(1) - 2i D_{\beta\beta\alpha\alpha,\uparrow\downarrow\uparrow\downarrow}(1)~.
\end{align}
\end{subequations}
Again, the SU(2) symmetry gives rise to the second equality of the third equation.
The above charge sum rules can be used to determine $\Lambda^{ch}$ self-consistently, provided that the generalized double occupancies are converged by the spin channel iterations.
\end{widetext}

It is worth noting that as the on-site inter-orbital hopping approaches zero, the off-diagonal elements of the density matrix vanish ($n_{\alpha\beta} = 0$ for $\alpha \neq \beta$), assuming that no spontaneous exciton condensate forms.
As a result, $\tilde{\Lambda}^{sp}$ in the pair-hopping sector diverges because the denominator in Eq.~\eqref{eq: renormalization factor} goes to zero, and TPSC fails in this limit.
To overcome this limitation, one can adopt the assumption $\tilde{\Lambda}^{sp}_{\alpha\alpha\beta\beta} = \tilde{\Lambda}^{sp}_{\alpha\beta\beta\alpha}$, as proposed in Refs. \cite{Zantout2019, Profe2025}.
Alternatively, formulating the multi-orbital TPSC in the rotated bonding/anti-bonding basis could provide a potential solution, as the on-site densities appear in the denominator in this case, helping to circumvent the singularity problem.
In this study, however, we focus on a bilayer Hubbard model and therefore employ the TPSC in the original basis, since the renormalized vertex in the pair-hopping channel, $\tilde{\Lambda}^{sp}_{\alpha\alpha\beta\beta}$, vanishes entirely.

\subsection{Spectral self-energy expression\label{subsec: self-energy expression}}

After self-consistently calculating the vertices on the two-particle level, the resulting corrections to the single-particle spectra can be incorporated through the spectral self-energy $\Sigma^{spc}$, which is obtained by solving the Swinger-Dyson equation (SDE) \cite{Stefanucci2013,Vilk1997b}.
It is important to distinguish $\Sigma^{spc}$ from the self-energy in Eq.~\eqref{eq: self-energy with HF Ansatz}; the former is obtained in a post-processing step, after the two-particle self-consistent calculations. 
Specifically, this is accomplished by considering Eqs.~\eqref{eq: equation of motion}, \eqref{eq: generalized susceptibility} and \eqref{eq: bethe-salpeter equation simplified}.
After some simplifications, we obtain $\Sigma^{spc} = \Sigma^{hf} + \Sigma^{cor}$, where
\begin{dmath}
\Sigma_{\alpha\beta}^{hf}(1,2) = \delta(1-2) U^{ch}_{\alpha\bar{\gamma}\beta\bar{\delta}}(1) n_{\bar{\gamma}\bar{\delta}}(1)
\end{dmath}
is the Hartree-Fock part with $n_{\beta\alpha}(1) = \avg{c^{\dag}_{\beta}(1) c_{\alpha}(1)}$.
The correlation self-energy reads
\begin{dmath}\label{eq: schwinger-dyson equation}
\Sigma_{\alpha\beta}^{cor}(1,2) = 
\frac{i}{8} G_{\bar{\delta}\bar{\epsilon}}(1,2) \left[ \tilde{\chi}_{\bar{\epsilon}\alpha\beta\bar{\delta}}^{ch}(2,1) + 3\tilde{\chi}_{\bar{\epsilon}\alpha\beta\bar{\delta}}^{sp}(2,1) \right]~,
\end{dmath}
where 
\begin{dmath}
\tilde{\chi}^{sp/ch}_{\alpha\beta\gamma\delta}(1,2) = \tilde{\Lambda}^{sp/ch}_{\alpha\bar{\beta}\gamma\bar{\delta}}(1) \chi^{sp/ch}_{\bar{\delta}\bar{\gamma}\bar{\gamma}\bar{\alpha}}(1,2) U^{sp/ch}_{\bar{\alpha}\beta\bar{\gamma}\delta}(2)~.
\end{dmath}
In the above formulas, we introduced
$
{U}^{ch}_{\alpha\beta\gamma\delta}(1) = 2{U}_{\alpha\beta\gamma\delta}(1) - {U}_{\alpha\beta\delta\gamma}(1)
$
and
$
{U}^{sp}_{\alpha\beta\gamma\delta}(1) = {U}_{\alpha\beta\delta\gamma}(1),
$
corresponding to the bare Coulomb tensors in the spin and charge channels, respectively.
Equation~\eqref{eq: schwinger-dyson equation} describes the influence of quantum fluctuations in the particle-hole (ph) channel on the single-particle spectrum.
If we set $\tilde{\Lambda}^{sp/ch}_{\alpha\beta\gamma\delta}(1) = U^{sp/ch}_{\alpha\beta\gamma\delta}(1)$, Eq.~\eqref{eq: schwinger-dyson equation} reduces to RPA, and its self-consistent solution is known as the fluctuation-exchange (FLEX) approach \cite{Vilk1997b}.

In this study, we employ the steady-state implementation of these formulas. 
Their derivation involves recasting the above contour formalism into the real-time Keldysh space by applying Langreth's rules \cite{Stefanucci2013}, followed by a Fourier transform to the momentum-frequency domain \cite{PhysRevB.105.085122,PhysRevB.109.155113}. 
This approach avoids numerically ill-defined analytical continuation from the Matsubara formalism, at the cost of using denser frequency meshes along the real axis.
The latter issue can be partially mitigated by introducing a fictitious bath with a small coupling constant $\Gamma$ to broaden the sharp peaks in the spectral functions \cite{PhysRevB.105.085122,Yan2023}.

\section{Numerical results\label{sec: numerical results}}

\begin{figure}
\includegraphics[width=1.0\linewidth]{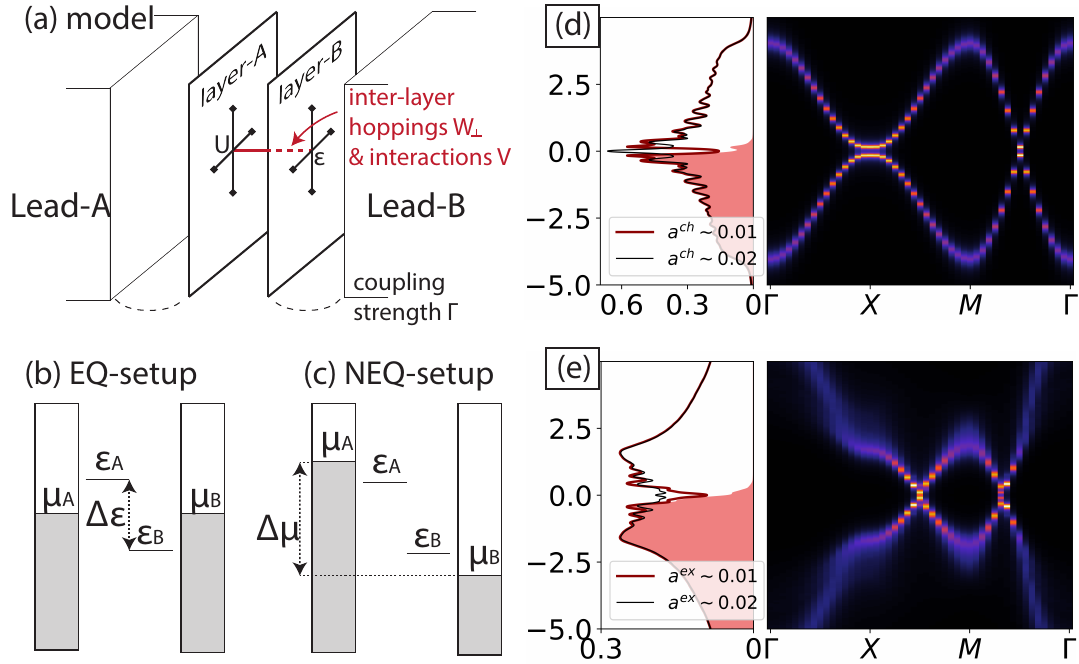}
\caption{(a) Illustration of the bi-layer structure coupled to two leads.
(b) and (c) depict the energy diagram of the equilibrium ($\mu_A - \mu_B = 0$) and the nonequilibrium ($\mu_A - \mu_B = \Delta\mu$) setups.
$\mu$ and $\epsilon$ refer to the chemical potential and the on-site energy of the lead and layer, respectively.
Panels (d) and (e) show the RPA results of the local and $\kvec$-resolved spectral functions in the critical region close to the charge (d) and excitonic (e) ordering, respectively.
The parameters are provided in the text.
\label{fig: model-PG}}
\end{figure}

We apply the multiorbital TPSC formalism to bi-layer stacks of the square lattice Hubbard model with both on-site interactions $U$ and interlayer repulsive interactions $V$, as illustrated in Fig.~\ref{fig: model-PG}(a).
In this case, the bare Coulomb tensor in Eq.~\eqref{eq: Hamiltonian} takes the specific form
$U_{i,\alpha\beta\gamma\delta} = U\delta_{\alpha\gamma}\delta_{\beta\delta}\delta_{\alpha\beta} + V \delta_{\alpha\gamma}\delta_{\beta\delta}(1-\delta_{\alpha\beta})$, 
i.e., we consider the two sites interacting by $V$ as the inter-orbital interactions in a two-orbital Hubbard description ($J = J^C = 0$ in Eq.~\eqref{eq: Kanamori-type interaction}). 
The indices $\alpha\cdots\delta \in \{A,B\}$ refer to the two layers, as shown in Fig.~\ref{fig: model-PG} (a). 
We consider only the nearest-neighbor (N.N.) hoppings between the lattices sites.
Thus the hopping matrix assumes the form $T_{ij,\alpha\beta} = \epsilon_\alpha \delta_{ij} \delta_{\alpha\beta} + W_\alpha \delta_{\avg{ij}} \delta_{\alpha\beta} + W_\perp \delta_{ij} (1-\delta_{\alpha\beta})$.
The in-plane N.N. hopping amplitude is used as the energy unit, i.e. $W_{A} = -W_{B} =1$.
Additionally, we introduce two non-interacting electron baths, each coupled to one of the layers, with the chemical potentials denoted by $\mu_{A/B}$.
These baths serve two purposes: (i) they drive the system out of equilibrium when their chemical potentials or temperatures differ, and (ii) they broaden the electronic states in the layers, facilitating rapid convergence of the real-frequency calculations.
In this manuscript, we assume a coupling strength of $\Gamma = 0.01$  (wide-band limit) between the baths and the layers \cite{PhysRevB.105.085122}.

In the bi-layer case, the bare Coulomb tensor in the spin and charge sector reads (see Appendix \ref{sec: orbital resolved BSE})
\begin{equation}\label{eq: spin and charge Coulomb tensor}
U^{sp} = 
\begin{pmatrix}
U & & & \\
&  V & & \\
& & V & \\
 & & & U
\end{pmatrix},~
U^{ch} = 
\begin{pmatrix}
U & & & 2V\\
& -V & & \\
& & -V & \\
2V & & & U
\end{pmatrix}.
\end{equation}
For the model without interactions ($U=V=0$), the dispersions of the $A$ and $B$ layers are opposite due to $W_A = -W_B = 1$, and their on-site energies are oppositely shifted from the Fermi level, i.e. $\epsilon_A - E_f = E_f - \epsilon_B = \Delta \epsilon /2$, to make them electron and hole doped.
When $U$ and $V$ (both positive) are introduced, electron correlations can induce spin, charge, exciton, or Cooper pair (not discussed here) fluctuations, which compete with each other.

\subsection{Order parameters and susceptibilities\label{subsec: order parameter}}

To quantify these different instabilities in an unbiased way, we calculate lattice susceptibilities in the normal phase (states without long-range orders). 
Only the susceptibilities in the ph-channel are discussed.
Because only local interactions (see Eq.~\eqref{eq: interacting Hamiltonian}) are considered in our model, we introduce a rotated local density as the order parameter \cite{Murakami2017,Murakami2020}
\begin{dmath}\label{eq: rotated local density matrix}
\rho^{\mu\nu}(1) = 
\hat{\sigma}^\mu_{\bar{\sigma}\bar{\sigma}'} \hat{\sigma}^\nu_{\bar{\alpha}\bar{\beta}} \avg{ c^\dag_{\bar{\alpha}\bar{\sigma}}(1) c_{\bar{\beta}\bar{\sigma}'}(1)}~, 
\end{dmath}
which allows to define the amplitude and phase mode of the excitons.
Here, $\hat{\sigma}^{i}$ ($i = 0,1,2,3$) is the $i$-th Pauli matrix.
The superscripts $\mu$ and $\nu$ correspond to indices in the spin and orbital (pseudo-spin) space, respectively.
One can easily show that $\hat{\rho}^{\mu\nu}(1)$ (without the expectation value) is a Hermitian operator, i.e. $[\hat{\rho}^{\mu\nu}(1)]^\dag = \hat{\rho}^{\mu\nu}(1)$.

\begin{table*}
\centering
\begin{tabular}{c|cccc}
\toprule
\diagbox{$\mu$ (spin)}{$\nu$ (orbital)} & $0$ (bonding) & $1$ (amplitude) & $2$ (phase) & $3$ (anti-bonding) \\
\midrule
$0$ (charge) & total charge & singlet exciton (amp.) & singlet exciton (pha.) & orbital polarization \\
$1$ (spin-$x$) & spin-$x$ density & triplet-$x$ exciton (amp.) & triplet-$x$ exciton (pha.) & orbital-selective spin-$x$ \\
$2$ (spin-$y$) & spin-$y$ density & triplet-$y$ exciton (amp.) & triplet-$y$ exciton (pha.) & orbital-selective spin-$y$ \\
$3$ (spin-$z$) & spin-$z$ density & triplet-$z$ exciton (amp.) & triplet-$z$ exciton (pha.) & orbital-selective spin-$z$ \\
\bottomrule
\end{tabular}
\caption{Physical interpretation of each component of the rotated local density matrix $\rho^{\mu\nu}$ defined in Eq.~\eqref{eq: rotated local density matrix}.}
\label{tab:rho-matrix-meaning}
\end{table*}

In this study, where the interlayer hoppings $W_\perp$ are assumed to be real, it follows from Eq.~\eqref{eq: rotated local density matrix} that $\rho^{\mu 1}$ and $\rho^{\mu 2}$ describe the amplitude ($\nu = 1$) and phase ($\nu = 2$) mode of the singlet ($\mu = 0$) and triplet ($\mu = 1,2,3$) exciton order parameters in the a system with $W_\perp > 0$.
(The amplitude and phase modes are indistinguishable when $W_\perp = 0$.)
Divergent susceptibilities in these channels indicate the onset of exciton condensation, corresponding to the spontaneous formation of particle-hole bound states. 

The phase mode characterizes fluctuations in the relative phase between the two orbitals (or bands) forming the exciton. In the condensed phase, for $W_\perp = 0$, this mode becomes gapless due to spontaneous U(1) symmetry breaking, and can be associated with the Goldstone mode of the excitonic condensate.
$\rho^{\mu 0}$ and $\rho^{\mu 3}$ yield the charge ($\mu = 0$) and spin ($\mu = 1,2,3$) order parameters defined in the bonding ($\nu = 0$) and anti-bonding ($\nu = 3$) orbital basis, respectively.
Physically, the $\nu=0$ component captures symmetric combinations of orbitals (i.e., uniform density/spin across the orbitals), while the $\nu=3$ component captures staggered orbital order, such as orbital polarization or orbital-selective magnetism.
Table \ref{tab:rho-matrix-meaning} summarizes the physical meaning of each component.

With the local order parameters defined in Eq.~\eqref{eq: rotated local density matrix}, one can introduce correlation functions that correlate two order parameters at different space-time point. This gives rise to the susceptibilites 
\begin{dmath}\label{eq: susceptibility}
i\chi^{\mu\nu, \mu'\nu'}(1,2) = \avg{\mathbb{T}\{\delta\hat{\rho}^{\mu\nu}(1), \delta\hat{\rho}^{\mu'\nu'}(2)\}}~,
\end{dmath}
where $\delta\hat{O}(1) = \hat{O} - \avg{\hat{O}}$.
From Eq.~\eqref{eq: generalized susceptibility}, one can easily obtain
\begin{dmath}
i\chi_{\alpha\beta\gamma\delta,\sigma_1\sigma_2\sigma_3\sigma_4}(1,2)
=
\avg{\mathbb{T}
\delta \hat{n}_{\gamma\alpha,\sigma_3\sigma_1}(1) \delta \hat{n}_{\delta\beta,\sigma_4\sigma_2}(2)
}~,
\end{dmath}
which leads to
\begin{dmath}\label{eq: susceptibility2}
\chi^{\mu\nu,\mu'\nu'}(1,2) = 
\hat{\sigma}^\mu_{\bar{\sigma}_2\bar{\sigma}_4}
\hat{\sigma}^{\mu'}_{\bar{\sigma}_1\bar{\sigma}_3}
\hat{\sigma}^{\nu}_{\bar{\beta}\bar{\delta}}
\hat{\sigma}^{\nu'}_{\bar{\alpha}\bar{\gamma}}
\chi_{\bar{\delta}\bar{\gamma}\bar{\beta}\bar{\alpha},\bar{\sigma}_4\bar{\sigma}_3\bar{\sigma}_2\bar{\sigma}_1}(1,2)~.
\end{dmath}
If the system exhibits spin SU(2) invariance, we have
$\chi^{0\nu,0\nu'} = \hat{\sigma}^\nu_{\bar{\beta}\bar{\delta}} \hat{\sigma}^{\nu'}_{\bar{\alpha}\bar{\gamma}} \chi^{ch}_{\bar{\delta}\bar{\gamma}\bar{\beta}\bar{\alpha}}$ and $\chi^{\mu\nu,\mu\nu'} = \hat{\sigma}^\nu_{\bar{\beta}\bar{\delta}} \hat{\sigma}^{\nu'}_{\bar{\alpha}\bar{\gamma}} \chi^{sp}_{\bar{\delta}\bar{\gamma}\bar{\beta}\bar{\alpha}} (\mu=1,2,3)$.

The bilayer model discussed in this work satisfies this condition.
Furthermore, the singlet and triplet exciton states are degenerate at $W_\perp = 0$, so that we omit singlet/triplet in this case.
In the case of more general Kanamori-type inter-orbital interactions, the degeneracy can be broken and lower the triplet exciton states \cite{PhysRevB.90.245144}.
Also for $W_\perp > 0$, we will only consider the spin-triplet exciton channel.

\subsection{RPA analysis of the equilibrium setup ($W_\perp=0$)\label{subsec: rpa}}

We first discuss the equilibrium setup of the bilayer model, as illustrated in Fig.~\ref{fig: model-PG}(a).
The interlayer hoppings $W_\perp$ are set to zero for the equilibrium study \footnote{For the equilibrium setup, the chemical potentials of the baths that coupled to the two respective layers are in equilibrium, i.e. $\mu_\alpha = \mu_\beta = 0$}.
As $U$ and $V$ increase, the disordered phase begins to develop short-range order.
Specifically, $U$ favors in-plane anti-ferromagnetic spin correlations at half-filling, while $V$ induces charge and exciton correlations, depending on the fillings of each band.
Long-range orders may form beyond the critical couplings where the corresponding susceptibilities diverge.
As seen from Eq.~\eqref{eq: bethe-salpeter equation simplified}, this critical condition is reached when the denominator matrix of the BSE becomes singular.
We thus introduce an `$a$' parameter~\cite{Janiifmmodecheckselsevsfi2008}
\begin{equation}
a =  1 + \varrho \left\{ \pm \frac{1}{2} \chi^0(\omega=0) \Lambda^{sp/ch} \right\}~,
\end{equation}
which defines the distance to this singularity.
Here, $\varrho\{{A}\}$ denotes the spectral radius, i.e. the maximum eigenvalue of the matrix ${A}$.
The disordered phase is stable when $a > 0$ and undergoes a transition to long-range order (LRO) when $a$ switches to negative.

Since in our case, there is no tunneling between the layers, $G_{\alpha\beta} \sim G_{\alpha\alpha} \delta_{\alpha\beta}$, and $\chi^0$ becomes diagonal in orbital space, i.e. $\chi^0 = \text{diag}(\chi^0_{AAAA}, \chi^0_{ABBA}, \chi^0_{BAAB}, \chi^0_{BBBB})$.
Since $\Lambda$ is assumed to have the same structure as the bare $U$ in Eq.~\eqref{eq: spin and charge Coulomb tensor}, the middle 2-by-2 submatrix of $\chi^0 \Lambda^{sp/ch}$ is decoupled from the rest.
Consequently, one can separate the spin-charge and the exciton sectors in solving BSEs.
Specifically, the $a$-parameters in these two sectors read
\begin{subequations}\label{eq: a-coefficient}
\begin{dmath}
a^{sp/ch} = 1 + \frac{1}{2}
\varrho 
\left\{\pm
\begin{bmatrix}
\chi^0_{AAAA}\Lambda^{sp/ch}_{AAAA} & \chi^0_{AAAA}\Lambda^{sp/ch}_{ABAB} \\
\chi^0_{BBBB}\Lambda^{sp/ch}_{BABA} & \chi^0_{BBBB}\Lambda^{sp/ch}_{BBBB}
\end{bmatrix}
\right\}~,
\end{dmath}
\begin{dmath}
a^{exsp/exch} = 1 + \frac{1}{2}
\varrho 
\left\{\pm
\begin{bmatrix}
\chi^0_{ABBA}\Lambda^{sp/ch}_{ABBA} & \\
& \chi^0_{BAAB}\Lambda^{sp/ch}_{BAAB}
\end{bmatrix}
\right\}~.
\end{dmath}
\end{subequations}
Here, $\chi^0$ is evaluated at zero-frequency, implying that only the static instability is considered in our equilibrium setup.
Since $\Lambda^{sp}_{\alpha\beta\beta\alpha} = -\Lambda^{ch}_{\alpha\beta\beta\alpha}$ ($\alpha\neq\beta$), we have $a^{exsp} = a^{exch} \equiv a^{ex}$, and hence a degeneracy of the singlet and triplet exciton states in the bi-layer system.
Note that the entire Brillouin zone must be explored to determine the $\qvec$-point that exhibits the most significant fluctuations.

We begin with an analytical study of the influence of the nonlocal interaction $V$ ($U = 0$) on the bi-layer model using the RPA approach, where $\tilde{\Lambda}^{sp/ch} = U^{sp/ch}$.
At half-filling, $\chi^0_{\alpha\alpha\alpha\alpha} = \chi^0_{\alpha\beta\beta\alpha}$ and since $U^{sp}_{\alpha\beta\beta\alpha} = U^{ch}_{\alpha\beta\alpha\beta}/2 = V$, we have $a^{ch} - a^{ex} = \chi^0(\omega=0) V/2 < 0$, i.e., the charge fluctuations dominate the excitonic fluctuations at half-filling.
The minimum of $a^{ch}$ occurs when $\qvec = (\pi,\pi)$ with eigenvector $\frac{1}{\sqrt{2}}(\ket{\alpha\alpha\alpha\alpha} - \ket{\beta\beta\beta\beta})$.
Physically, this means that the charge correlations exhibit a checkerboard pattern within each layer and are staggered between the layers.
When $a^{ch} \rightarrow 0$, the spectral function, calculated using the RPA $\chi$ in Eq.~\eqref{eq: schwinger-dyson equation}, exhibits the opening of a pseudo-gap, as shown in Fig.~\ref{fig: model-PG}(d).
Here, the black and red lines refer to $a^{ch} \approx 0.02$ and $a^{ch} \approx 0.01$, respectively. (Parameters: $U=0$, $V=0.834$, $\Delta \epsilon = 0$, $\Gamma = 0.05$, half-filling;  $\beta = 9$ for $a^{ch} \approx 0.02$ and $\beta = 10$ for $a^{ch} \approx 0.01$.)
The right panel plots the corresponding $\kvec$-resolved band structure when $a^{ch} \approx 0.01$.
Physically, the opening of the pseudo-gap at the Fermi level when $a^{ch}$ is reduced from $0.02$ to $0.01$, is due to the 
growing collective fluctuations that scatter electrons and partially suppress the spectral weight.

A crystal field splitting $\Delta \epsilon$ (Fig.~\ref{fig: model-PG}(b)) quickly suppresses the charge fluctuations in the system, resulting in dominant excitonic fluctuations.
Figure~\ref{fig: model-PG}(e) shows the spectral function (red) and its $\kvec$-resolved structure for $a^{ex} \approx 0.01$. (Parameters: $U=0$, $V=2.6$, $\Delta \epsilon = 1.0$, $\Gamma = 0.05$. The system is half-filled with $\beta = 9$ for $a^{ex} \approx 0.02$ and $\beta = 10$ for $a^{ex} \approx 0.01$.) 
Compared to $a^{ex} \approx 0.02$ (black), we observe a similar pseudo-gap opening, which in this case originates from the strong excitonic fluctuations \cite{PhysRevLett.120.216401}.

\begin{figure}
\includegraphics[width=1.0\linewidth]{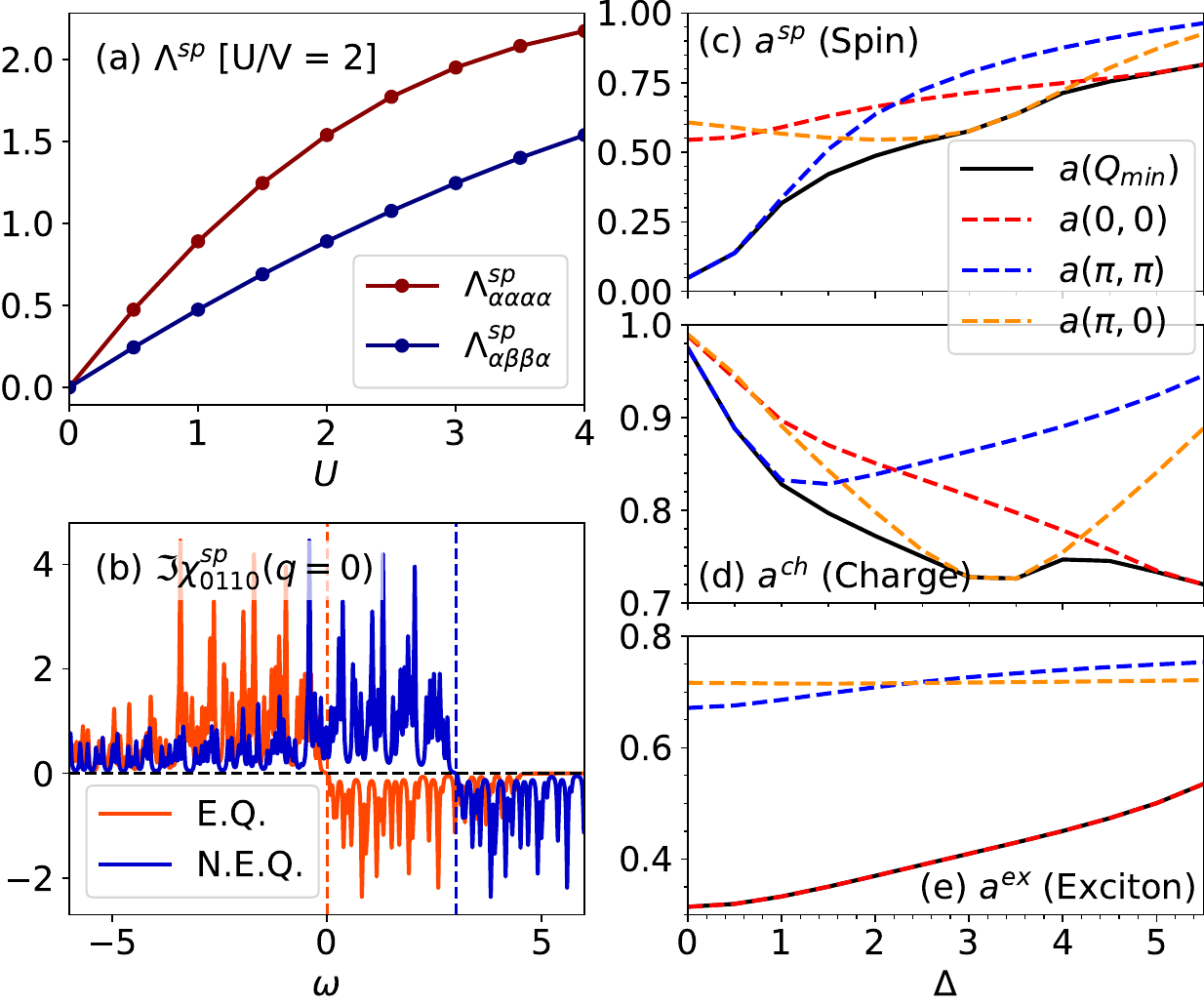}
\caption{(a) Spin vertex $\Lambda^{sp}$ vs on-site Coulomb interaction $U$. (b) Imaginary part of the exciton susceptibility $\chi^{sp}_{ABBA}(\qvec = 0, \omega)$ with red and blue lines corresponding to the EQ and NEQ models, respectively. (c), (d), (e): Spin, charge and exciton instabilities $a$ as a function of crystal field splitting $\Delta \epsilon$. The black solid line shows the global minimum across the 1BZ, while dashed lines show the results for high symmetry points. Parameters: $U=4.0$, $V=2.0$ and $\beta = 4.0$.\label{fig: eq-data}
}
\end{figure}

\subsection{TPSC results for the equilibrium setup ($W_\perp=0$)\label{subsec: tpsc for wperp = 0}}

We next consider a more realistic case with both nonzero $U$ and $V$.
Instead of using RPA, which can quickly lead to a divergence in the spin channel, we apply the TPSC approach, which renormalises both the spin and charge vertices using the local spin and charge sum rules.
In Fig.~\ref{fig: eq-data}(a), we plot the renormalized vertices $\Lambda^{sp/ch}$ as a function of their bare values. (Calculation performed at half-filling with $U/V = 2$.)
One can see that both $\Lambda^{sp}_{AAAA}$ and $\Lambda^{sp}_{ABBA}$ saturate at large $U$, known as Kanamori-Brueckner screening.
This fact prevents the system from developing magnetic and excitonic long-range orders, consistent with the Mermin-Wagner theorem.

In the absence of interlayer tunneling, introducing a crystal field splitting $\Delta \epsilon$ in the equilibrium (EQ) model is equivalent to shifting the chemical potential by $\Delta \mu := \mu_A - \mu_B = \Delta \epsilon$, which we refer to in the following as the nonequilibrium (NEQ) model.
The EQ and NEQ models are related by a Gauge transformation (see Appendix \ref{sec: gauge transform}), and their susceptibilities satisfy the relation
\begin{dmath}
\chi^{\text{NEQ}}_{\alpha\beta\beta\alpha}(\omega,\qvec) = \chi^{\text{EQ}}_{\alpha\beta\beta\alpha}(\omega - \bar{\delta}_{\alpha\beta} \Delta \mu,\qvec)~,
\end{dmath}
where $\bar{\delta}_{\alpha\beta} = 1 - \delta_{\alpha\beta}$.
In Fig.~\ref{fig: eq-data} (b), we plot the imaginary part of $\chi^{sp}_{ABBA}(\omega; \qvec=0)$ for both the EQ ($\Delta \epsilon = -3$, $\Delta \mu = 0$) and NEQ ($\Delta \mu = 3$, $\Delta \epsilon = 0$) models, which shows that in the NEQ case, $\chi$ is exactly shifted on the $\omega$ axis by $\Delta \mu$.
Moreover, it can be rigorously shown that $\Im \chi$ vanishes at $\omega = 0$ in the EQ model, which implies that a static instability ($\omega = 0$) can occur in the EQ setup.
This static instability in the EQ model corresponds to a dynamical instability in the NEQ model, namely to a state with rotating phase of the order parameter \cite{Perfetto2019,Murakami2020a,Osterkorn2025}.

In Fig.~\ref{fig: eq-data}(c-e), we illustrate the static instabilities 
in the spin, charge and exciton sectors of the EQ model by plotting the $a$ parameters as a function of $\Delta \epsilon$.
The black solid lines represent the maximum instability (minimum value of $a$) identified in a search over the whole BZ.
Additionally, the results for high-symmetry points are shown by the dashed lines.
At $\Delta \epsilon= 0$, the smallest value of $a^{sp}$ occurs at $(\pi,\pi)$, indicating that AFM spin fluctuations dominate the system.
As $\Delta \epsilon$ increases, spin fluctuations are rapidly suppressed, as is seen by the quick increase in $a^{sp}$ in panel (c).
In contrast, charge fluctuations are enhanced, as evidenced by the decreasing $a^c$ curves in panel (d).
Interestingly, the momentum associated with these instabilities, $Q_{min}$, undergoes an incommensurate evolution, shifting along $(\pi,\pi) \rightarrow (\pi,0) \rightarrow (0,0)$ in both the spin and charge channels.
This evolution results in a non-trivial increase (decrease) in the spin (charge) susceptibility.

\begin{figure}
\includegraphics[width=1.0\linewidth]{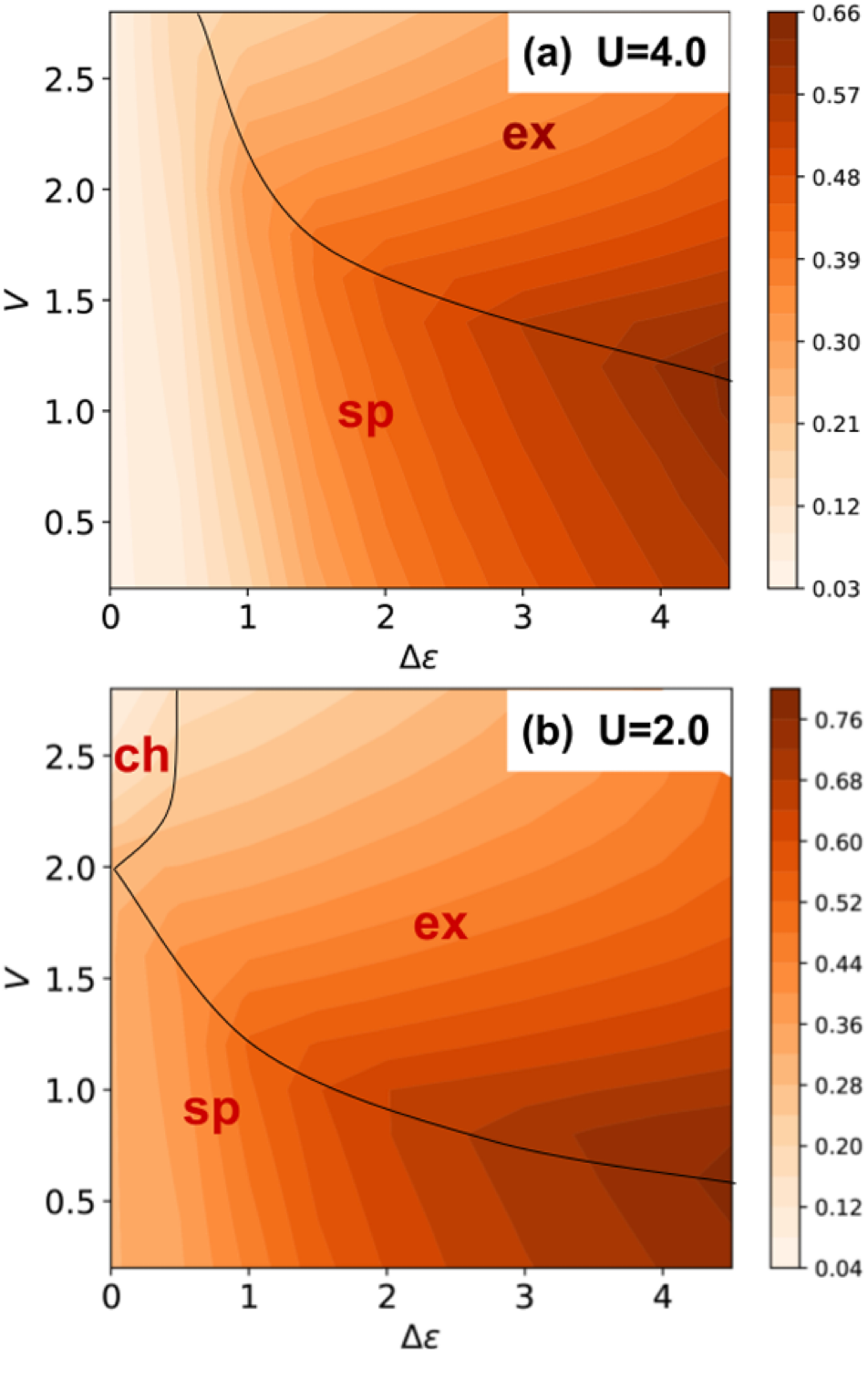}
\caption{Phase diagram of the bilayer model in the $V$-$\Delta\epsilon$ plane for 
(a) $U=4.0$ and (b) $U=2.0$. The color map represents the minimum value of the 
$a$-coefficient across different sectors. Parameters: $W_\perp=0.0$ and $\beta=4.0$.
\label{fig: eq-phasediagram}}
\end{figure}

For the exciton sector, as shown in panel (e), $a^{ex}$ increases 
monotonically with increasing $\Delta \epsilon$ with $Q_{min}$ fixed at $(0,0)$. 
This shows that excitons are formed by the electrons and holes with the same momentum $\kvec$, as expected for our setup with opposite dispersions of the two bands, and that $\Delta\epsilon>0$ suppresses exciton formation. 
Comparing the values for the different sectors in Fig.~\ref{fig: eq-data}(c-e), we see that the system transitions from an AFM-dominated regime to an EI dominated regime with increasing $\Delta \epsilon$, which is consistent with the fRG results reported in Ref.~\cite{Scherer2018}.

To examine phase instabilities across the parameter space, Fig.~\ref{fig: eq-phasediagram} shows the minimum $a$-coefficients for the sectors defined in Eq.~\eqref{eq: a-coefficient}, with $\beta=4.0$.
Panels (a) and (b) correspond to $U=4.0$ and $U=2.0$, respectively. 
For small $V$ and $\Delta\epsilon$, the on-site $U$-driven AFM fluctuations at $\qvec=(\pi,\pi)$ dominate, as seen in the lower-left corner of both panels. 
Increasing $\Delta\epsilon$ rapidly suppresses the spin fluctuations, allowing excitonic fluctuations to take over. 
When $V>U$ and $\Delta\epsilon$ is small, a charge-sector instability emerges, consistent with the analytical discussion in Sec.~\ref{subsec: rpa}. 
For large $V$ and $\Delta\epsilon$, we further expect a strong charge instability at $\qvec=(0,0)$ (not shown), as suggested by Fig.~\ref{fig: eq-data}, where the charge-sector instability grows monotonically 
with $\Delta\epsilon$, while the other channels weaken.

\subsection{TPSC results for the model with interlayer hopping ($W_\perp>0$)\label{subsec: tpsc for wperp != 0}}

We now examine the influence of the interlayer hopping $W_\perp$ and the bias voltage $\Delta\mu = \mu_A - \mu_B$ between the two layers (Fig.~\ref{fig: model-PG}(b,c)) on the excitonic properties.
This setup is directly related to engineered bilayer heterostructures of two-dimensional materials, where the control of interlayer excitons and their condensates are actively explored \cite{Xie2018,Ma2021}.

Unlike the case $W_\perp = 0$, where $\Delta \mu$ can always be gauge-transformed into an effective shift $\Delta \epsilon’$ applied to the bilayer system, this equivalence no longer holds for $W_\perp > 0$.
Specifically, particle number conservation within each layer is broken, which breaks the U(1) gauge symmetry.
Consequently, the amplitude and phase modes become distinct.
From a mathematical perspective, $W_\perp > 0$ makes $\chi^0$ non-diagonal in orbital space, inducing a mixing between the triplet-spin and singlet-charge channels.
We focus on the triplet-spin channel, which hosts the dominant quantum fluctuations in our bilayer model \cite{PhysRevB.90.245144}.

\subsubsection{Equilibrium setup}

\begin{figure*}
\includegraphics[width=1.0\linewidth]{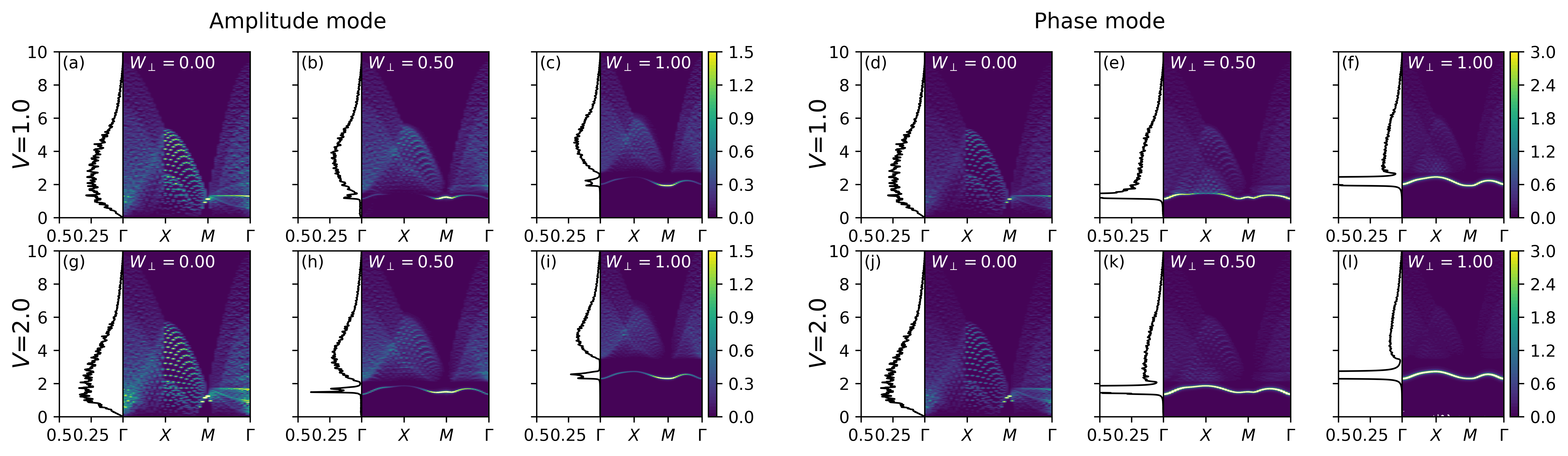}
\caption{Spectra of the exciton susceptibility, $-\Im \chi^{sp}_{\text{amp/pha}}(\omega,\qvec)/\pi$, showing the amplitude and phase modes (Eq.~\eqref{eq: susceptibility2}) for different interlayer hoppings, $W_\perp = 0.0,,0.5,,1.0$. The top and bottom rows correspond to $V=1.0$ and $V=2.0$, respectively. The parameters used are $U=4$, $\Delta\epsilon=1.5$ and $\Delta\mu = 0$.\label{fig: eq-spc}}
\end{figure*}

We first consider the equilibrium case with $\Delta \mu = 0$.
The on-site energies of the two layers are shifted by $\Delta \epsilon = \epsilon_A - \epsilon_B$.
Figure \ref{fig: eq-spc} shows the spectra of $\chi^{sp}$ in both the amplitude and phase channels, calculated according to Eq.~\eqref{eq: susceptibility2} and denoted by $\chi^{sp}_{\text{amp/pha}}$, respectively.
The results are obtained for $U = 4$ and $\Delta \epsilon = 1.5$, with two values of the interlayer interaction: $V = 1$ (top row) and $V = 2$ (bottom row). 
In each panel, the left plot presents the $\kvec$-summed local spectrum, while the right plot displays the dispersion along the high-symmetry path in the first Brillouin zone: $\Gamma=(0,0) \to X=(\pi,0) \to M=(\pi,\pi) \to \Gamma=(0,0)$.
The calculations are performed in the disordered state. 
For $W_\perp=0$, the amplitude and phase spectra are identical [panels (a,d) and (g,j)] due to the SU(2) pseudo-spin symmetry [Eq.~\eqref{eq: rotated local density matrix}].
An interlayer hopping $W_\perp>0$ (assumed to be real) induces hybridization between the two layers, breaking the U(1) symmetry and serving as an order parameter.
The resulting phase is an insulating state, which can be regarded as a forced excitonic insulator when strong excitonic correlations develop even in the absence of $W_\perp$.
Due to the reduced symmetry, the collective modes get gapped.
As seen in panels (d-f), a gapped mode emerges with increasing $W_\perp$ and exhibits hardening.
Note that at nonzero $\qvec$ and/or with nonzero $W_\perp$ the phase and amplitude channels are mixed so that one can see the same collective in-gap signals in both channels.
Still, the intensity of the signal is stronger in the phase channel, which suggest a predominant phase mode character of the mode.
Increasing $V$ further sharpens this feature: from panel (e) to (k), the gapped phase mode changes from weak and continuum-like to sharp and well defined, signaling enhanced excitonic bound states.

\subsubsection{Nonequilibrium setup}

We now switch to the nonequilibrium properties of the model by applying a voltage bias between the baths that couple to the respective layers, see Fig.~\ref{fig: model-PG}(c).
This bias drives the system into a nonequilibrium steady state.
In the following, we fix the interaction strengths to $U = 4.0$ and $V = 2.0$, and set the on-site energy difference of the two layers to $\Delta \epsilon = 2.1$.
The interlayer hopping is fixed at $W_\perp = 1$, a regime where the exciton bound state becomes particularly pronounced.

\begin{figure}
\includegraphics[width=1.0\linewidth]{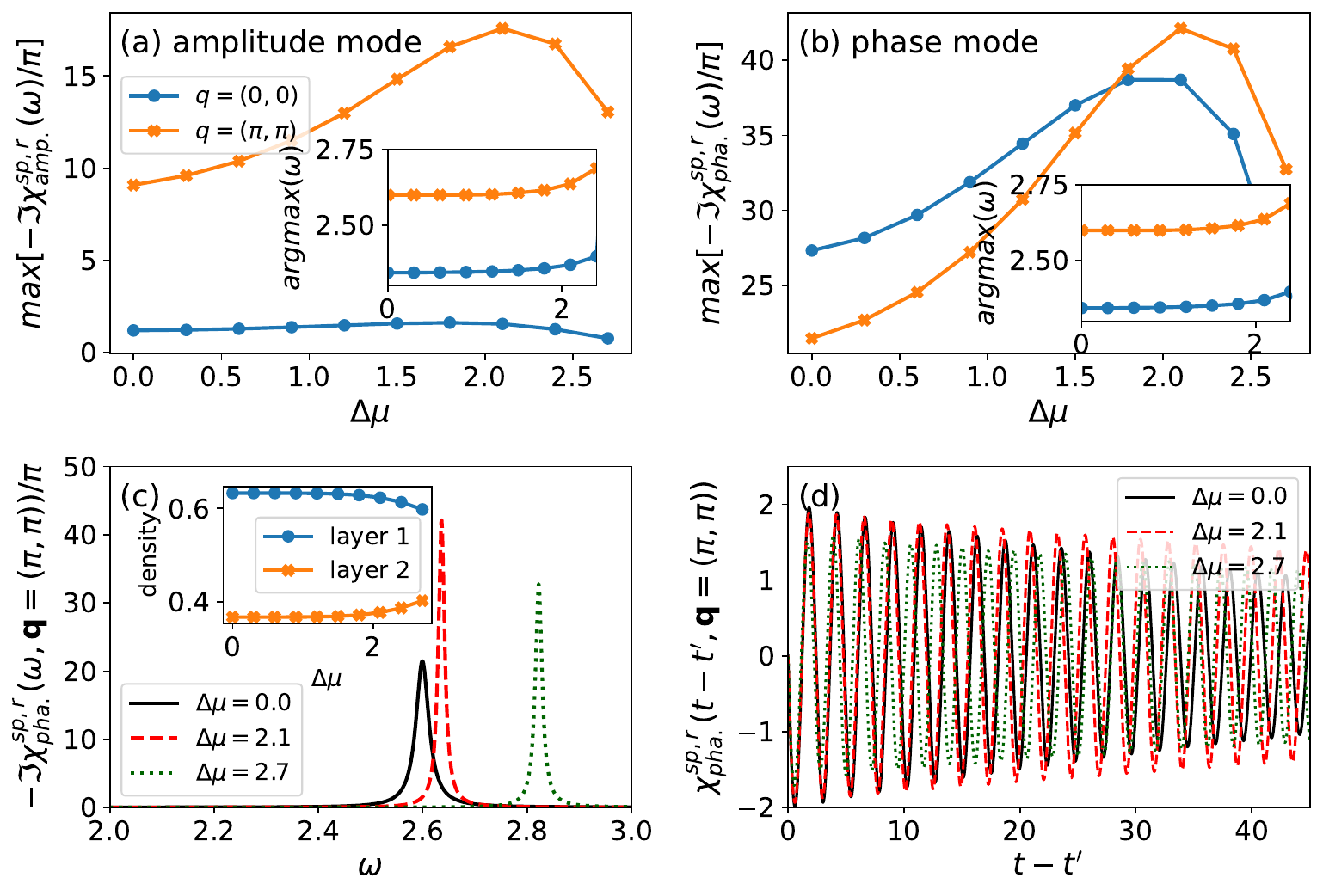}
\caption{
(a),(b) Maximum value of $-\Im\chi^{sp,r}(\omega)/\pi$ across all frequencies for the amplitude and phase modes. The insets show the corresponding frequency positions of these maxima.
(c) The exciton susceptibility for the phase mode $-\Im\chi^{sp,r}(\omega, \qvec = (\pi,\pi))/\pi$ for $\Delta \mu = 0.0$ (black), $\Delta \mu = 2.1$ (red dashed) and $\Delta \mu = 2.7$ (green dotted).
Inset: densities vs $\Delta \mu$ for each layer.
(d) Time evolution of the susceptibility $\chi^{sp,r}(t-t', \qvec=(\pi,\pi))$, which is real.
\label{fig: NEQ-exciton-mode}}
\end{figure}

Figures~\ref{fig: NEQ-exciton-mode}(a) and \ref{fig: NEQ-exciton-mode}(b) show the maximum value of the excitonic susceptibility spectra, i.e. $\max[-\Im\chi^{sp,r}_{\text{amp/pha}}(\omega,\qvec)/\pi]$, 
for the amplitude and phase channels at two momenta: $\qvec=(0,0)$ (in blue) and $\qvec=(\pi,\pi)$ (in orange).
Here, the ``$r$" in the superscript denotes the retarded component. 
With increasing bias $\Delta \mu$, the excitonic bound states in both the amplitude and phase mode are initially enhanced, demonstrating a bias-induced amplification of the excitonic mode. 
The maximum value of $-\Im \chi$ occurs when $\Delta \mu \approx \Delta \epsilon$.
In the rotated frame, this condition corresponds to $\Delta \mu = \Delta \epsilon = 0$, but with a time-dependent interlayer hopping $e^{i\Delta \mu t}$, see Appendix \ref{sec: gauge transform}.
The enhanced excitonic bound states are thus primarily related to the van Hove singularity.
The insets of panel (a) and (b) show the peak frequencies  that are shifted slightly upward with bias, meaning that the exciton excitation energy is modified by the nonequilibrium setup.
Upon further increasing the applied bias, $\Delta \mu > \Delta \epsilon$, the peak rapidly diminishes due to the charge imbalance between the two layers, as can be seen in the inset of panel (c).
Furthermore, the crossing of the blue and orange curves in Fig.~\ref{fig: NEQ-exciton-mode}(b) indicates that the phase excitations at $\qvec=(0,0)$ and $\qvec=(\pi,\pi)$ become nearly degenerate.
Note that, even when $\Delta \mu \approx \Delta \epsilon$, the phase mode does not diverge, since the nonzero interlayer hopping $W_\perp$ explicitly breaks the U(1) symmetry in each layer.

To further clarify the spectral signatures of the bias induced modifications, Fig.~\ref{fig: NEQ-exciton-mode}(c) plots the frequency dependence of the phase mode spectrum $-\Im\chi^{sp,r}_{\text{pha}}(\omega,\qvec=(\pi,\pi))/\pi$.
In equilibrium ($\Delta\mu = 0$, black solid line), one observes an excitonic peak located near $\omega \simeq 2.6$. 
Under a bias of $\Delta \mu = 2.1$ (red dashed line), the peak becomes significantly enhanced and sharper, indicating a weaker damping of the mode. 
For an even larger bias of $\Delta \mu = 2.7$ (green dotted line), the peak gets suppressed and broadened, signaling increasingly detrimental effects of the nonequilibrium state on the excitonic susceptibility. 
To reveal the origin of this behavior, the inset of Fig.~\ref{fig: NEQ-exciton-mode}(c) displays the electron densities of the two layers as a function of the applied bias $\Delta \mu$.
The densities remain close to their equilibrium values until the local chemical potential crosses the on-site layer energy at $\Delta \epsilon = 2.1$. Beyond this point, there is a strong charge redistribution, and the resulting imbalance suppresses the excitonic peak.

Finally, Fig.~\ref{fig: NEQ-exciton-mode}(d) provides the real time perspective by plotting the phase-mode susceptibility $\chi^{sp,r}_{\text{pha}}(t-t’,\qvec=(\pi,\pi))$.
In equilibrium ($\Delta\mu=0$), the response displays coherent oscillations with a slow decay, representative of a well-defined but damped exciton mode.
When the bias is tuned into the resonant regime ($\Delta\mu=2.1$), the oscillations persist for a longer time, indicating a longer-lived exciton.
This directly illustrates the stabilization of interlayer coherence under nonequilibrium driving.

\section{Conclusions\label{sec: conclusions}}

In this work, we presented our nonequilibrium extension of the multi-orbital TPSC method.
We employed the recently proposed Hartree–Fock ansatz to decouple the equations of motion, which provides improvements over the traditional Hartree ansatz, particularly in the regime of small Hund’s coupling.
The vertex corrections in the spin and charge channels are computed self-consistently at the two-particle level by using the local spin and charge sum rules.
The screening of spin vertices suppresses the spurious finite-temperature transitions that plague DMFT and many-body perturbative methods in two dimensions.
For the multi-orbital case, additional spin and charge sum rules consistent with SU(2) symmetries are imposed to close the vertex self-consistency loop.
By formulating the theory within the Keldysh nonequilibrium Green’s function framework, our approach works directly on the real-frequency axis, thereby eliminating the need for numerically ill-defined analytic continuation.
It thus produces reliable spectral functions with well-resolved fine structures.
The trade-off is that much more frequency points are required compared to the Matsubara formulation.

As an application, we applied the nonequilibrium TPSC formalism to the bilayer Hubbard model to investigate correlation effects for various interaction strengths, interlayer tunnelings, and bias conditions.
We showed that sufficiently strong fluctuations in the spin, charge, or excitonic channels can open a pseudogap near the Fermi level, indicating precursor behavior to long range ordered phases.
We systematically examined instabilities in the normal phase with respect to the onsite and interlayer Coulomb interactions $U$ and $V$, as well as the crystal field splitting between the two layers $\Delta\epsilon$.
Our results revealed that the system exhibits incommensurate ordering tendencies 
(in both the spin and charge channels) between two commensurate phases as $\Delta\epsilon$ increases.
The corresponding phase diagram, calculated in the disordered regime, is shown in Fig.~\ref{fig: eq-phasediagram}, highlighting the competition between spin, charge, and excitonic fluctuations across the parameter space of the model.

To study the effect of nonequilibrium electron populations, we applied a voltage bias across the bilayer by shifting the local chemical potentials of the leads coupled to each layer.
We computed the excitonic spectral function and clearly identified gapped excitonic bound states in the presence of interlayer tunneling.
These bound states are significantly enhanced under a nonzero interlayer bias, but are rapidly suppressed for voltage biases larger than $\Delta\epsilon$, 
due to the growing charge imbalance between the layers.

A future combination of our  nonequilibrium multi-orbital TPSC with density functional theory input will enable realistic simulations of low-dimensional correlated materials.

\begin{acknowledgments}
We thank Roser Valenti for valuable discussions.
The calculations were performed on the Beo06 cluster at the University of Fribourg and the SIMYC cluster at the Shanghai Institute of Microsystem and Information Technology (SIMIT).
P.W. and J.Y. acknowledge support from SNSF Grant No. 200021-196966.
J.Y. also acknowledges the startup fund from SIMIT, the National Natural Science Foundation of China (Grant No. 12504288) and the Science and Technology Commission of Shanghai Municipality (Grant No. 25ZR1402550).
J. B. Profe acknowledge support by the Deutsche Forschungsgemeinschaft (DFG, German Research Foundation) for funding through project QUAST-FOR5249 - 449872909 (project TP4).
Y. M. acknowledge support from Grant-in-Aid for Scientific Research from JSPS, KAKENHI Grant Nos. JP21H05017, JP24H00191 and JP25K07235.
\end{acknowledgments}

\appendix

\section{Orbital-resolved Bethe-Salpeter equations\label{sec: orbital resolved BSE}}

The Bethe-Salpeter equations for the spin and charge channels, expressed in the orbital space, read
\begin{dmath}\label{eq: multi-orbital bse in ph-channel}
\chi^{sp/ch}_{\alpha\beta\gamma\delta}(1,2) = \chi^0_{\alpha\beta\gamma\delta}(1,2) \mp \frac{1}{2} \chi^0_{\alpha\bar{\gamma}\gamma\bar{\alpha}}(1,\bar{1}) \Lambda^{sp/ch}_{\bar{\alpha}\bar{\beta}\bar{\gamma}\bar{\delta}}(\bar{1}) \chi^{sp/ch}_{\bar{\delta}\beta\bar{\beta}\delta}(\bar{1},2)~,
\end{dmath}
where 
\begin{equation}
\chi^0_{\alpha\beta\gamma\delta}(1,2) = 
-2iG_{\alpha\delta}(1,2^+)G_{\beta\gamma}(2,1^+)
\end{equation}
is the bare electron-hole bubble.
In the particle-hole channel, one can group $a:= (\alpha\gamma)$ and $b:=(\delta\beta)$, and Eq.~\eqref{eq: multi-orbital bse in ph-channel} can be recast into the matrix equations
\begin{dmath}
\chi^{sp/ch}_{ab}(1,2) = \chi^0_{ab}(1,2) \mp \frac{1}{2}\chi^0_{a\bar{a}}(1,\bar{1})\Lambda^{sp/ch}_{\bar{a}\bar{b}}(\bar{1})\chi^{sp/ch}_{\bar{b}b}(\bar{1},2)~.
\end{dmath}

In the description of the two-orbital model, we denote the two orbitals by $A$ and $B$.
We use the convention $a := (AA,AB,BA,BB)$ and similarly for $b$.
As a result, the Bethe-Salpeter equations become
\begin{widetext}
\begin{small}
\begin{dmath}
\begin{pmatrix}
\chi^{sp/ch}_{AAAA} & \chi^{sp/ch}_{ABAA} & \chi^{sp/ch}_{AAAB} & \chi^{sp/ch}_{ABAB}\\
\chi^{sp/ch}_{AABA} & \chi^{sp/ch}_{ABBA} & \chi^{sp/ch}_{AABB} & \chi^{sp/ch}_{ABBB}\\
\chi^{sp/ch}_{BAAA} & \chi^{sp/ch}_{BBAA} & \chi^{sp/ch}_{BAAB} & \chi^{sp/ch}_{BBAB}\\
\chi^{sp/ch}_{BABA} & \chi^{sp/ch}_{BBBA} & \chi^{sp/ch}_{BABB} & \chi^{sp/ch}_{BBBB}
\end{pmatrix}(1,2)
=
\begin{pmatrix}
\chi^{0}_{AAAA} & \chi^{0}_{ABAA} & \chi^{0}_{AAAB} & \chi^{0}_{ABAB}\\
\chi^{0}_{AABA} & \chi^{0}_{ABBA} & \chi^{0}_{AABB} & \chi^{0}_{ABBB}\\
\chi^{0}_{BAAA} & \chi^{0}_{BBAA} & \chi^{0}_{BAAB} & \chi^{0}_{BBAB}\\
\chi^{0}_{BABA} & \chi^{0}_{BBBA} & \chi^{0}_{BABB} & \chi^{0}_{BBBB}
\end{pmatrix}(1,2)
\mp
\frac{1}{2}
\begin{pmatrix}
\chi^{0}_{AAAA} & \chi^{0}_{ABAA} & \chi^{0}_{AAAB} & \chi^{0}_{ABAB}\\
\chi^{0}_{AABA} & \chi^{0}_{ABBA} & \chi^{0}_{AABB} & \chi^{0}_{ABBB}\\
\chi^{0}_{BAAA} & \chi^{0}_{BBAA} & \chi^{0}_{BAAB} & \chi^{0}_{BBAB}\\
\chi^{0}_{BABA} & \chi^{0}_{BBBA} & \chi^{0}_{BABB} & \chi^{0}_{BBBB}
\end{pmatrix}(1,\bar{1})
\begin{pmatrix}
\Lambda^{sp/ch}_{AAAA} &  &  & \Lambda^{sp/ch}_{ABAB}\\
 & \Lambda^{sp/ch}_{ABBA} & \Lambda^{sp/ch}_{AABB} & \\
 & \Lambda^{sp/ch}_{BBAA} & \Lambda^{sp/ch}_{BAAB} & \\
\Lambda^{sp/ch}_{BABA} &  &  & \Lambda^{sp/ch}_{BBBB}
\end{pmatrix}(\bar{1})
\begin{pmatrix}
\chi^{sp/ch}_{AAAA} & \chi^{sp/ch}_{ABAA} & \chi^{sp/ch}_{AAAB} & \chi^{sp/ch}_{ABAB}\\
\chi^{sp/ch}_{AABA} & \chi^{sp/ch}_{ABBA} & \chi^{sp/ch}_{AABB} & \chi^{sp/ch}_{ABBB}\\
\chi^{sp/ch}_{BAAA} & \chi^{sp/ch}_{BBAA} & \chi^{sp/ch}_{BAAB} & \chi^{sp/ch}_{BBAB}\\
\chi^{sp/ch}_{BABA} & \chi^{sp/ch}_{BBBA} & \chi^{sp/ch}_{BABB} & \chi^{sp/ch}_{BBBB}
\end{pmatrix}(\bar{1},2) .
\end{dmath}
\end{small}
The bare Kanamori-type interactions (Eq.~\eqref{eq: Kanamori-type interaction}) in this basis read
\begin{equation}
U^{sp}(1) = 
\begin{pmatrix}
U_{AA} &  &  & U_{AB}\\
 & J_{AB} & J^{C}_{AB} & \\
 & J^{C}_{BA} & J_{BA} & \\
U_{BA} &  &  & U_{BB}
\end{pmatrix}(1)~,
\quad
U^{ch}(1) = 
\begin{pmatrix}
2U_{AA} &  &  & 2U_{AB} - J_{AB}\\
 & 2J_{AB} - U_{AB} & J^{C}_{AB} & \\
 & J^{C}_{BA} & 2J_{BA} - U_{BA} & \\
2U_{BA} - J_{BA} &  &  & 2U_{BB}
\end{pmatrix}(1)~.
\end{equation}
\end{widetext}

\section{Gauge transformation of the bi-layer Hubbard model\label{sec: gauge transform}}

This section discusses the gauge transformation of the bi-layer Hubbard model, with each layer coupled to a non-interacting bath.
For simplicity, we omit the electron-electron interactions, which are irrelevant to the present discussion.
The Hamiltonian of the model then reads
\begin{dmath}
\Hop(t) = \sum_{\alpha}\left[ \Hop^{ld}_{\alpha}(t) + \Hop^{hyb}_{\alpha} \right] + \Hop^{lyr}(t)~,
\end{dmath}
where
\begin{subequations}
\begin{align}
\Hop^{ld}_{\alpha}(t) &= \sum_{mn}H^{ld}_{mn} a^{\dag}_{\alpha m} a_{\alpha n} + \mu_\alpha(t) \sum_{m} \hat{n}_{\alpha m}^a ,\\
\Hop^{hyb}_{\alpha} &= \sum_{im} \left[ H^{hyb}_{im} c_{i\alpha}^\dag a_{m\alpha} + h.c. \right] ,\\
\Hop^{lyr}(t) &= \sum_{\alpha,ij} H^{lyr}_{\alpha,ij} c_{i\alpha}^\dag c_{j\alpha} + \sum_{\alpha} \epsilon_\alpha(t) \sum_i \hat{n}_{\alpha i}^c \nonumber\\ 
&+ \sum_i \left[ W_\perp c^\dag_{iA} c_{iB} + h.c. \right].
\end{align}
\end{subequations}
Here, $m$, $n$ ($i$, $j$) are site indices in the baths (layers).
$\hat{n}_{\alpha m}^a = a_{\alpha m \bar{\sigma}}^\dag a_{\alpha m \bar{\sigma}}$ and $\hat{n}_{\alpha i}^c = c_{\alpha i \bar{\sigma}}^\dag c_{\alpha i \bar{\sigma}}$ are the occupation number operators of $a$- and $c$-electrons in the baths and layers, respectively.
$\mu_\alpha(t)$ refers to a homogeneous voltage applied to the $\alpha$-lead, and $\epsilon_\alpha(t)$ represents a time-dependent on-site energy shift in the bi-layer structure.

To simplify the discussions and reveal the underlying physics, we introduce a unitary transformation to the wave function $\ket{\tilde{\Psi}(t)} = \Uop(t) \ket{\Psi(t)}$, which satisfies the Schr\"odinger equation with the transformed Hamiltonian
\begin{dmath}
\tilde{\Hop}(t) = \Uop(t) \left[ \Hop(t) - i\frac{\partial}{\partial t} \right] \Uop^\dag(t)~.
\end{dmath}
In our case, we choose
\begin{dmath}
\Uop(t) = e^{i\int_{t_0}^t d\bar{t} \sum_{\alpha m} \mu_\alpha(\bar{t}) \hat{n}^a_{\alpha m}} e^{i\int_{t_0}^t d\bar{t} \sum_{\alpha i} \mu_\alpha(\bar{t}) \hat{n}^c_{\alpha i}}~.
\end{dmath}
After applying the Baker-Campbell-Hausdorff formula, we have
\begin{dmath}
\tilde{\Hop}(t) = \sum_{\alpha}\left[ \tilde{\Hop}^{ld}_{\alpha}(t) + \tilde{\Hop}^{hyb}_{\alpha} \right] + \tilde{\Hop}^{lyr}(t)~,
\end{dmath}
where
\begin{subequations}
\begin{align}
\tilde{\Hop}^{ld}_{\alpha} &= \sum_{mn}H^{ld}_{mn} a^{\dag}_{\alpha m} a_{\alpha n}~,\\
\tilde{\Hop}^{hyb}_{\alpha} &= \sum_{im} \left[ H^{hyb}_{im} c_{i\alpha}^\dag a_{m\alpha} + h.c. \right] ,\\
\tilde{\Hop}^{lyr}(t) &= \sum_{\alpha,ij} H^{lyr}_{\alpha,ij} c_{i\alpha}^\dag c_{j\alpha} + \sum_{\alpha} \left[ \epsilon_\alpha(t) - \mu_\alpha(t) \right] \sum_i \hat{n}_{\alpha i}^c \nonumber\\ 
&+ \sum_i \left[ W_\perp e^{i\int_{t_0}^t d\bar{t} \left[ \mu_A(\bar{t}) - \mu_{B}(\bar{t}) \right]} c^\dag_{iA} c_{iB} + h.c. \right].
\end{align}
\end{subequations}
From these expressions, it is evident that the time-dependence of the lead Hamiltonian is transferred to the layer Hamiltonian.
The relative on-site energy difference between the layer and its attached lead is fixed after the transformation.
At the same time, an additional time-dependent phase is introduced to $W_\perp$,
which can be interpreted as a perpendicular electric field across the bi-layer structure.

In this work, we only focus on the time-independent steady-state solution, which can be viewed as the low-frequency limit of the above model.
When $W_\perp = 0$, it is obvious that there is no modulation of the interlayer tunnelings, and in this case the NEQ model  is equivalent to the EQ model, see Fig.~\ref{fig: model-PG}(b,c).
The two gauges introduced above, in principle, give the same physics.
However, the (off-diagonal) Green's function as well as the susceptibilities, in general, are gauge dependent.
Their relationship satisfies
\begin{dmath}
\avg{\hat{S}^z_{i,\alpha\beta}}_{eq}(t) = \avg{\hat{S}^z_{i,\alpha\beta}}_{neq}(t) \mathbb{T} e^{-i\int_{t_0}^t d\bar{t} \mu_\alpha(\bar{t}) } \mathbb{T} e^{i\int_{t_0}^t d\bar{t} \mu_\beta(\bar{t})}~.
\end{dmath}
If we fix the bias $\mu_{A/B}(t)$ to a time-independent constant, i.e. $\mu_{A} = - \mu_{B} = \Delta \mu/2$, the above expression becomes $\avg{\hat{S}^z_{i,AB}}_{eq}(t) = \avg{\hat{S}^z_{i,AB}}_{neq}(t) e^{-i \Delta\mu (t-t_0) }$.
The off-diagonal elements rotate, corresponding to a rotating exciton order parameter.
The correlation function satisfies the relationship
\begin{equation}
\avg{\hat{S}^z_{i,AB} \hat{S}^z_{j,BA}}_{eq}(t - t') = \avg{\hat{S}^z_{i,AB} \hat{S}^z_{j,BA}}_{neq}(t - t') e^{-i \Delta\mu (t-t')}~,
\end{equation}
which in the frequency domain reads
\begin{dmath}
\avg{\hat{S}^z_{i,AB} \hat{S}^z_{j,BA}}_{eq}(\omega + \Delta \mu) = \avg{\hat{S}^z_{i,AB} \hat{S}^z_{j,BA}}_{neq}(\omega)~.
\end{dmath}

\bibliography{ref}

\end{document}